\documentclass{pasj00}

\begin{document}
\SetRunningHead{K. Yabe et al.}{The Mass-Metallicity Relation of Star-Forming Galaxies at $z\sim1.4$}

\title{NIR Spectroscopy of Star-Forming Galaxies at $z\sim1.4$ with Subaru/FMOS: The Mass-Metallicity Relation}

\author{Kiyoto \textsc{Yabe}\altaffilmark{1}, Kouji \textsc{Ohta}\altaffilmark{1}, Fumihide \textsc{Iwamuro}\altaffilmark{1}, Suraphong \textsc{Yuma}\altaffilmark{1}, Masayuki \textsc{Akiyama}\altaffilmark{2}, Naoyuki \textsc{Tamura}\altaffilmark{3}, Masahiko \textsc{Kimura}\altaffilmark{3}, Naruhisa \textsc{Takato}\altaffilmark{3}, Yuuki \textsc{Moritani}\altaffilmark{1}, Masanao \textsc{Sumiyoshi}\altaffilmark{1}, Toshinori \textsc{Maihara}\altaffilmark{1}, John \textsc{Silverman}\altaffilmark{4}, Gavin \textsc{Dalton}\altaffilmark{5,6}, Ian \textsc{Lewis}\altaffilmark{5}, David \textsc{Bonfield}\altaffilmark{5,7}, Hanshin \textsc{Lee}\altaffilmark{5,8}, Emma \textsc{Curtis Lake}\altaffilmark{5,9}, Edward \textsc{Macaulay}\altaffilmark{5}, and Fraser \textsc{Clarke}\altaffilmark{5}}

\altaffiltext{1}{Department of Astronomy, Kyoto University, Sakyo-ku, Kyoto, 606-8502, Japan}
\email{kiyoyabe@kusastro.kyoto-u.ac.jp}
\email{ohta@kusastro.kyoto-u.ac.jp}
\altaffiltext{2}{Astronomical Institute, Tohoku University, Aoba-ku, Sendai, 980-8578, Japan}
\altaffiltext{3}{Subaru Telescope, National Astronomical Observatory of Japan, 650 North A'ohoku Place, Hilo, HI 96720, USA}
\altaffiltext{4}{Institute for the Physics and Mathematics of the Universe, The University of Tokyo, Kashiwanoha, Kashiwa, 277-8583, Japan}
\altaffiltext{5}{Department of Astrophysics, University of Oxford, Keble Road, Oxford OX1 3RH, UK}
\altaffiltext{6}{STFC Rutherford Appleton Laboratory, Chilton, Didcot, Oxfordshire OX11 0QX, UK}
\altaffiltext{7}{Centre for Astrophysics Research, Science and Technology Research Institute, University of Hertfordshire, Hatfield AL10 9AB, UK}
\altaffiltext{8}{McDonald Observatory, University of Texas at Austin, 1 University Station C1402, Austin, TX 78712, USA}
\altaffiltext{9}{Institute for Astronomy, University of Edinburgh, Royal Observatory, Edinburgh EH9 3HJ, UK}


%


\maketitle

\begin{abstract}
We present near-infrared spectroscopic observations of star-forming galaxies at $z\sim1.4$ with FMOS on the Subaru Telescope. We observed K-band selected galaxies in the SXDS/UDS fields with $K\leq23.9$ mag, $1.2\leq z_{ph} \leq 1.6$, $M_{*}\geq 10^{9.5} M_{\odot}$, and expected F(H$\alpha$) $\geq$ 10$^{-16}$ erg s$^{-1}$ cm$^{-2}$. 71 objects in the sample have significant detections of H$\alpha$. For these objects, excluding  possible AGNs identified from the BPT diagram, gas-phase metallicities are obtained from [N\emissiontype{II}]/H$\alpha$ line ratio. The sample is split into three stellar mass bins, and the spectra are stacked in each stellar mass bin. The mass-metallicity relation obtained at $z\sim1.4$ is located between those at $z\sim0.8$ and $z\sim2.2$. We constrain an intrinsic scatter to be $\sim0.1$ dex or larger in the mass-metallicity relation at $z\sim1.4$; the scatter may be larger at higher redshifts. We found trends that the deviation from the mass-metallicity relation depends on the SFR and the half light radius: Galaxies with higher SFR and larger half light radii show lower metallicities at a given stellar mass. One possible scenario for the trends is the infall of pristine gas accreted from IGM or through merger events. Our data points show larger scatter than the fundamental metallicity relation (FMR) at $z\sim0.1$ and the average metallicities slightly deviate from the FMR. The compilation of the mass-metallicity relations at $z\sim3$ to $z\sim0.1$ shows that they evolve smoothly from $z\sim3$ to $z\sim0$ without changing the shape so much except for the massive part at $z\sim0$.
\end{abstract}

\section{Introduction}
The gas-phase metallicity (hereafter, metallicity) is a fundamental parameter for the understanding of formation and evolution of galaxies, because it traces past star formation activity; metals in gas are produced in stars formed and returned into the inter-stellar medium (ISM) of galaxies. The metallicity is also affected by inflow and outflow of gas. Thus, by investigating the metallicity and its cosmological evolution, the star formation history of galaxies together with the gas inflow and outflow can be constrained.

The presence of a correlation between stellar mass and metallicity (hereafter, mass-metallicity relation) in the local universe is well known. \citet{Tremonti:2004p4119} established the mass-metallicity relation at $z\sim0.1$ with a large sample ($\sim53,000$) of  Sloan Digital Sky Survey (SDSS) galaxies. Tracing the cosmological evolution of the mass-metallicity relation is indispensable to reveal how galaxies have been evolving. At $z\sim1$, the mass-metallicity relations and the downsizing-like evolution to $z\sim0.1$ were found; the less evolution can be seen in the massive part (e.g., \cite{Savaglio:2005p3325,Zahid:2011p11939}). However, the anti-downsizing-like evolution of the mass-metallicity relations were also presented at similar redshifts \citep{Lamareille:2009p5295, PerezMontero:2009p5308}. The mass-metallicity relations were obtained at $z\sim2$ by using $\sim9
0$ galaxies \citep{Erb:2006p4143} and at $z\sim3$ by $\sim20$ galaxies \citep{Maiolino:2008p5212, Mannucci:2009p8028} and the smooth evolution of the mass-metallicity relation from $z\sim3$ to $z\sim0$ is suggested. However, at $z\sim2$, higher metallicities at a given stellar mass have been reported \citep{Hayashi:2009p4235,Yoshikawa:2010p4286,Onodera:2010p4273} than those found by \citet{Erb:2006p4143}. The discrepancy at $z\sim2$ may be partly due to the small size of their samples ($10-20$). Since the redshift of $z=1-2$ is close to the peak epoch in the cosmic star-formation history, establishing the mass-metallicity relation at this redshift is very important.

Although the metallicity correlates well with stellar mass,  there is a scatter in the relation. Its origin may provide a clue to understand the process of the chemical enrichment and may be related to how the mass-metallicity relation evolves. At $z\sim0.1$, \citet{Tremonti:2004p4119} found that the mass-metallicity relation has an intrinsic scatter of $\sim0.1$ and the scatter is larger at the lower stellar mass. At a given stellar mass, galaxies with higher surface stellar mass density tends to show higher metallicity, suggesting that they transformed more gas into stars raising the metallicity. \citet{Ellison:2008p7997} showed that galaxies with larger specific star-formation rate (SFR) and size show lower metallicity at a given stellar mass at $z\sim0.1$. More recently, \citet{Mannucci:2010p8026} and \citet{Yates:2011p16030} found that galaxies with larger SFRs tend to show lower metallicities at $z\sim0.1$ and the scatter around the mass-metallicity relation is reduced very much (i.e., the fundamental metallicity relation) by introducing SFR as the second parameter affecting the metallicity. The scatter of the mass-metallicity relation, however, is still unclear at higher redshifts because of the limited sample size. In order to investigate these trends further, including the dependency of other parameters, at higher redshift, a large near-infrared spectroscopic survey is necessary, which is very time consuming and hard to achieve.

The Fiber Multi Object Spectrograph (FMOS) on the Subaru Telescope enables us to survey a large spectroscopic sample at high redshifts.  FMOS is a near infrared (NIR) fiber multi-spectrograph \citep{Kimura:2010p11396}.   The fiber positioner ``Echidna'' at the prime focus feeds 400 fibers in a FoV of \timeform{30'} diameter. Two NIR spectrographs (IRS1 and IRS2) with an OH airglow suppression system are capable of obtaining both low resolution ($R\sim650$) and high resolution ($R\sim3000$) spectra in the wavelength range of $0.9-1.8$$\mu$m. Taking advantage of the multiplicity available with FMOS, we are conducting a large spectroscopic survey for star-forming galaxies at $z=1-2$. This redshift range is of much interest, because it is near or shortly after the peak epoch in the cosmic star formation history (e.g., \cite{Hopkins:2006p4539}) and thus important for the history of the chemical evolution of galaxies. It is also suitable for FMOS, because in the redshift range [N\emissiontype{II}]$\lambda\lambda$6548,6584, H$\alpha$, [O\emissiontype{III}]$\lambda\lambda$4959,5007, and H$\beta$ lines are located in the wavelength range observable with FMOS. In this paper we present the first results from this survey as to metallicity measurements of galaxies at $z\sim1.4$.

Throughout this paper, we adopt the concordance cosmology, ($\Omega_{M}$ , $\Omega_{\Lambda}$ , $h$) = (0.3, 0.7, 0.7). All magnitudes are in the AB system \citep{Oke:1983p15127}.

\section{Sample and Observations}
\subsection{Photometric Data in the SXDS/UDS Fields\label{Sec_Data}}
The sample is selected from a galaxy catalog in the SXDS/UDS fields, where deep multi-wavelength data are available. The optical \textit{Subaru}/Suprime-Cam images ($B$, $V$, $R_{C}$, $i'$, and $z'$) are available from the Subaru-XMM Deep Survey (SXDS; \cite{Furusawa:2008p15159}) with the original pixel scale of \timeform{0''.202} pix$^{-1}$. The limiting magnitudes are 27.7, 27.2, 27.1, 27.0, and 26.0 mag for $B$, $V$, $R_{C}$, $i'$, and $z'$-band, respectively ($5\sigma$, \timeform{2''.0} diameter). The NIR \textit{UKIRT}/WFCAM images ($J$, $H$, and $K_{s}$) are taken from DR8 version of the UKIDSS Ultra Deep Survey (UDS; \cite{Lawrence:2007p15348}) with the original pixel scale of \timeform{0''.268} pix$^{-1}$. The limiting magnitudes are 24.9, 24.2, and 24.6 mag for $J$, $H$, and $K_{s}$, respectively ($5\sigma$, \timeform{2''.0} diameter). The \textit{Spitzer}/IRAC images ($3.6\mu$m, $4.5\mu$m, $5.8\mu$m, and $8.0\mu$m) are taken from the \textit{Spitzer} public legacy survey of the UKIDSS Ultra Deep Survey (SpUDS; PI: J. Dunlop, in preparation) with the original pixel scale of \timeform{0''.600} pix$^{-1}$. The limiting magnitudes are 24.8, 24.5, 22.8, and 22.7 mag for ch1 ($3.6\mu$m), ch2 ($4.5\mu$m), ch3 ($5.8\mu$m), ch4 ($8.0\mu$m), respectively ($5\sigma$, \timeform{2''.4} diameter). We use an overlapped area of these data. The effective area is $\sim0.67$ deg$^{2}$.

The NIR and MIR images are aligned to the optical images with pixel scale of \timeform{0''.202} pix$^{-1}$. Object detection and photometry in each image is carried out with respect to the reference $K_{s}$-band image by using the \textit{double image} mode of SExtractor \citep{Bertin:1996p15563}. For the optical and NIR images, aperture photometry with an aperture of \timeform{2''.0} diameter is carried out after applying the PSF matching. The total magnitude is calculated from the aperture magnitude scaled to the SExtractor's MAG\_AUTO in $K_{s}$-band. For the IRAC images, the total magnitude is calculated from the aperture magnitude with \timeform{2''.4} diameter applying aperture corrections. The correction factors are determined by generating artificial objects with various intrinsic sizes on the IRAC images and recovering the aperture magnitudes. The uncertainties of the aperture correction in the above simulations are 0.05, 0.05, 0.09, and 0.08 mag for ch1, ch2, ch3, and ch4, respectively.

\begin{figure*}
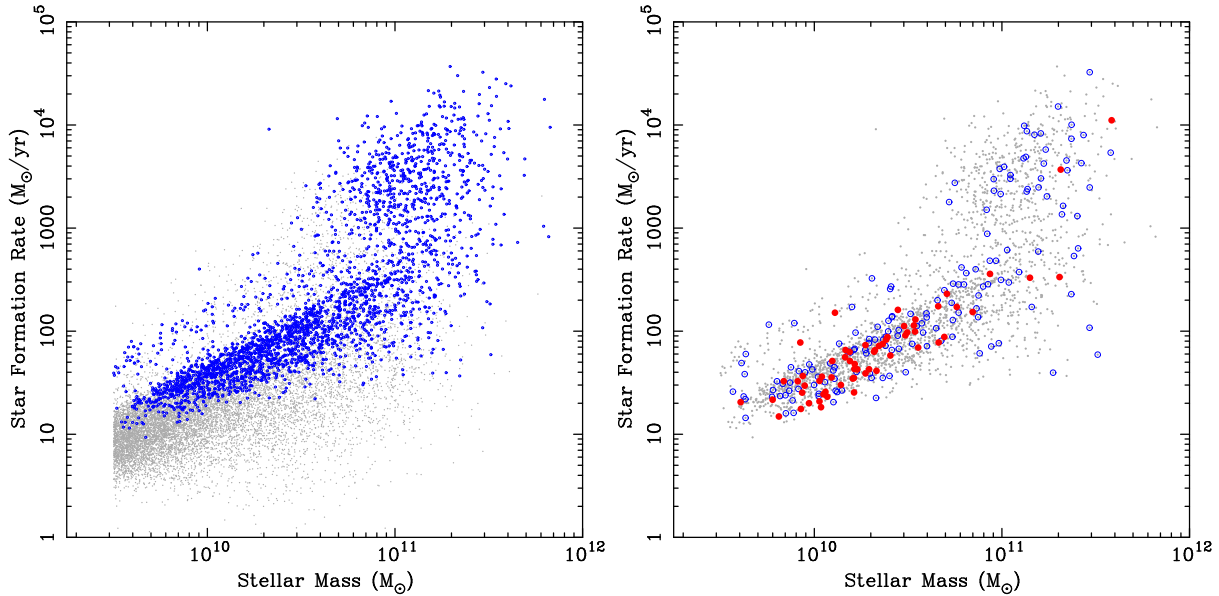

\begin{center}
\includegraphics[scale=0.45,angle=270]{f1a.eps}
\includegraphics[scale=0.45,angle=270]{f1b.eps}
\caption{Our targets on the $M_{*}$$-$SFR diagram. \textit{Left}: Our primary sample with $K_{s}<23.9$ mag, $1.2\leq z_{ph}\leq 1.6$, and $M_{*}\geq10^{9.5}M_{\odot}$ is indicated by \textit{gray dots}. Among the primary sample, the secondary sample with expected F(H$\alpha$)$\geq$1$\times$10$^{-16}$ erg s$^{-1}$ cm$^{-2}$ is indicated by \textit{blue open circles}. \textit{Right}: Objects observed with FMOS are plotted as \textit{circles}. While objects whose H$\alpha$ lines are detected with SN of $>$3 are indicated by \textit{red filled circles}, objects whose H$\alpha$ is not detected with SN of $>$3 are indicated by \textit{blue open circles}. The secondary sample are also plotted by \textit{gray dots}. \label{Fig_msfr}}
\end{center}
\end{figure*}

\subsection{Photometric Redshift Sample\label{Sec_Sample}}
For each source, the photometric redshift (photo-$z$) is determined by using \textit{Hyperz} \citep{Bolzonella:2000p243}, where model templates are fitted to the observed Spectral Energy Distributions (SEDs) from optical to MIR by a standard $\chi^{2}$ minimization. We use a standard set with synthetic spectra with 51 age grids from 0.3 Myr to 20 Gyr and 21 reddening ($A_{V}$) grids from 0.0 to 4.0 with $\Delta A_{V}=0.2$ mag, and the CWW \citep{Coleman:1980p17673} templates as well. The range of the photo-$z$ is set to be $0.0<z_{ph}<6.0$ with a step of $\Delta z$=0.03. The obtained photo-$z$ is compared with the available spectroscopic redshift (spec-$z$) in the SXDS/UDS (Simpson et al 2011, in preparation; Akiyama et al. 2011, in preparation; \cite{Smail:2008p15567}). Although some catastrophic outliers exist, the photo-$z$ generally agrees with the spec-$z$ with uncertainty of $\sigma_{z}/(1+z)=0.02-0.03$ at $z=1-2$.

In order to derive stellar masses of these galaxies, we perform SED fitting by using the \textit{SEDfit} \citep{Sawicki:2011}. For model SEDs, we use the BC03 population synthesis code \citep{Bruzual:2003p6157} with Padova 1994 evolutionary tracks. The Salpeter IMF \citep{Salpeter:1955p16612} with the mass range of 0.1$-$100 $M_{\odot}$ and the metallicity of 1.0 Z$_{\odot}$ are assumed. The extinction curve of \citet{Calzetti:2000p7012} is also assumed. We fix redshifts to those determined by the \textit{Hyperz} in order to reduce the number of free parameters. We generate model SEDs with 25 age grids from 0.2 Myr to 15.8 Gyr, 5 star formation histories (constant SF, $\tau=1$Gyr, $\tau=100$Myr, $\tau=10$Myr, and instantaneous burst) grids, and 21 $E(B-V)$ grids from 0.0 to 1.0 with a step of $\Delta E(B-V)=0.05$ mag. We then fit the model SEDs to the observed SEDs with the standard $\chi^{2}$ minimization.

The color excesses are estimated from the rest-frame UV colors. It is known that the rest-frame UV color strongly correlates with $E(B-V)$ \citep{Meurer:1999p6853}. For objects at $z\sim2$, \citet{Daddi:2004p6750,Daddi:2007p1460} presented the relation between $B-z'$ color and $E(B-V)$ by calibrating with the results from the SED fitting. For the redshift range of our sample, $B-z'$ color is not a good indicator of the extinction because the $z'$-band covers beyond the Balmer/4000\AA break. For $1.2<z<1.6$, we calibrate the relation between the $B-i'$ color and $E(B-V)$ by using the results of the SED fitting. The resulting relation is $E(B-V)=0.31(B-i')+0.02$. All the color excess of our sample is derived from the $B-i'$ color by using this relation. As we mention in Section 3, both H$\alpha$ and H$\beta$ lines are detected in some galaxies. For these objects, the color excess can be derived based on the Balmer decrement, by assuming  the intrinsic H$\alpha$$/$H$\beta$ ratio of 2.86 and the Calzetti extinction law. The average result is consistent with the color excess from the rest-frame UV colors within the error, if the extinction for the nebular emission is assumed to be larger than that for stellar continuum as we describe below.

The SFR is derived from the rest-frame UV luminosity density by using the conversion by \citet{Kennicutt:1998p7465}. The rest-frame UV luminosity density is calculated from the $B$-band magnitude. The intrinsic SFR is derived by correcting for the extinction with the $E(B-V)$ derived above. The Calzetti extinction curve is also assumed. For some objects, we cross-check with the SFRs from MIPS fluxes by using archived images of the SpUDS. We use the conversion from the rest-frame $L_{8\mu\textrm{m}}$ to the SFR by \citet{Daddi:2007p1460}. The resulting SFR from IR luminosity roughly agrees with that from the rest-frame UV luminosity density.

The expected H$\alpha$ flux is calculated from the intrinsic SFR and the $E(B-V)$ described above. Since it is suggested that the extinction is significantly larger for the ionized gas than for the stellar component \citep{Calzetti:2000p7012}, we convert the obtained $E(B-V)$ to that for the ionized gas by using a prescription by \citet{CidFernandes:2005p3252}. For the conversion from the SFR to the H$\alpha$ luminosity, we use the relation by \citet{Kennicutt:1998p7465}.

The sample is cross-correlated with X-ray sources \citep{Ueda:2008p13177} and objects which are cross-matched within the error circle of the X-ray source are excluded from the sample. Accordingly, the X-ray bright AGNs (L$_{X(2-10\textrm{keV})}$$\gtrsim$10$^{43}$ erg s$^{-1}$) are excluded from our sample.

\subsection{Sample\label{Sec_Selection}}
For the spectroscopic observations with FMOS, the primary sample is constructed from the K-selected catalogue described above with the following selection criteria: $K_{s}<23.9$ mag, $1.2\leq z_{ph}\leq 1.6$, and $M_{*}\geq 10^{9.5} M_{\odot}$. This primary sample size is very large, and many of the sample galaxies may show very faint H$\alpha$ emission lines. Thus in order to make an efficient survey, and as a first step, we selected objects for which  expected H$\alpha$ flux is larger than 1.0$\times$10$^{-16}$ erg s$^{-1}$ cm$^{-2}$ as the secondary sample. This selection could introduce some bias for the sample,  hence the effects will be considered later based on the results obtained.

In Figure \ref{Fig_msfr}, intrinsic SFRs derived from the rest-frame UV are plotted against the stellar mass (hereafter $M_{*}-SFR$ diagram) for the primary sample. The SFR correlates with the stellar mass. On the $M_{*}$$-$SFR diagram, the secondary sample is also shown with blue open circles. The distribution is slightly shifted toward larger SFR at a given stellar mass particularly in less massive part. This is due to the fact that we selected objects with larger expected H$\alpha$ flux and thus larger intrinsic SFR. We will discuss the effect of the selection bias on the metallicity in Section \ref{Sec_Bias}. Actually observed objects are randomly selected from the secondary sample. The right panel of Figure \ref{Fig_msfr} shows the observed objects among the secondary sample.

Some target objects are relatively massive ($M_{*}\gtrsim5\times 10^{10}M_{\odot}$) and have high intrinsic SFR ($\gtrsim 1000 M_{\odot} \textrm{yr}^{-1}$) as seen in Figure \ref{Fig_msfr}. We cross-correlated to the MIPS sources, and we found that the detection rate in MIPS is significantly lower for these objects than for other objects. The estimation of these high SFRs may be incorrect partly because the $E(B-V)$ is overestimated due to the dust/age degeneracy. We included these objects in our targeted sample and found that the actual detection rate of H$\alpha$ emission for the galaxies in this $M_{*}-SFR$ region is significantly lower than that in the other region as seen in right panel of Figure \ref{Fig_msfr}.

\subsection{Observations and Data Reduction\label{Sec_Obs}}
The observations were carried out with FMOS \citep{Kimura:2010p11396} on the Subaru Telescope on September $27-28$, November $11-13$, $21-24$, 28, and December $15-16$, 2010 (mostly the guaranteed time observing runs and partly engineering observing run for science verification, and open use observations). The observations were made with the Cross Beam Switch (CBS) mode in low resolution (LR) mode employing both IRS1 and IRS2. The spectral resolution in the LR mode is $R\sim500$ at $\lambda\sim1.10\mu m$, $R\sim650$ at $\lambda\sim1.30\mu m$, and $R\sim800$ at $\lambda\sim1.55\mu m$, which are measured from the Th-Ar lamp frames.The pixel scale in the LR-mode is 5 \AA. Our targets are allocated to fibers together with other scientific targets. Although weather conditions varied during the observing runs, the typical exposure time was $\sim3$ hours for one FMOS field of view. The typical seeing size measured with the FMOS sky-camera during the observations was \timeform{0''.9}  in $R$-band.

The obtained data are reduced with the FMOS pipeline \textit{FIBRE-pac} and detailed descriptions for the data reduction are presented by \citet{Iwamuro:2011}. Here we describe an outline of the process. For a set of exposures, the $A-B$ sky subtraction is carried out. Then the obtained 2D spectra are corrected for the distortion and residual sky subtraction  are also carried out. The obtained spectra are combined for the total exposures. Since the spectrum of one target is obtained by two fibers in the CBS mode, the final spectrum of one target is obtained by merging the CBS pair spectra. The wavelength calibration is done by using Th-Ar lamp frames. The uncertainty associated with the wavelength calibration is $\Delta \lambda \sim 5 \textrm{\AA}$. The relative flux calibration was done by using several F, G or K-type stars selected based on $J-H$ and $H-K_{s}$ colors and observed simultaneously with other scientific targets. The uncertainty of the relative flux calibration is  $\sim10\%$ at $\lambda=1.0-1.8\mu$m. The absolute flux is determined from the observed count rate by assuming the total throughput of the instrument, which is calibrated by using the moderately bright stars ($\sim$16 mag in $J$- or $H$-band) in the best weather conditions ($\sim$ \timeform{0''.7}) in previous engineering observations. The uncertainty of the absolute flux calibration itself is estimated to be $\sim 10\%$. However, there may be additional uncertainties ($20-30\%$) that come from variation of the atmospheric transmission and the seeing conditions.

\begin{figure*}
\begin{center}
\includegraphics[scale=0.35]{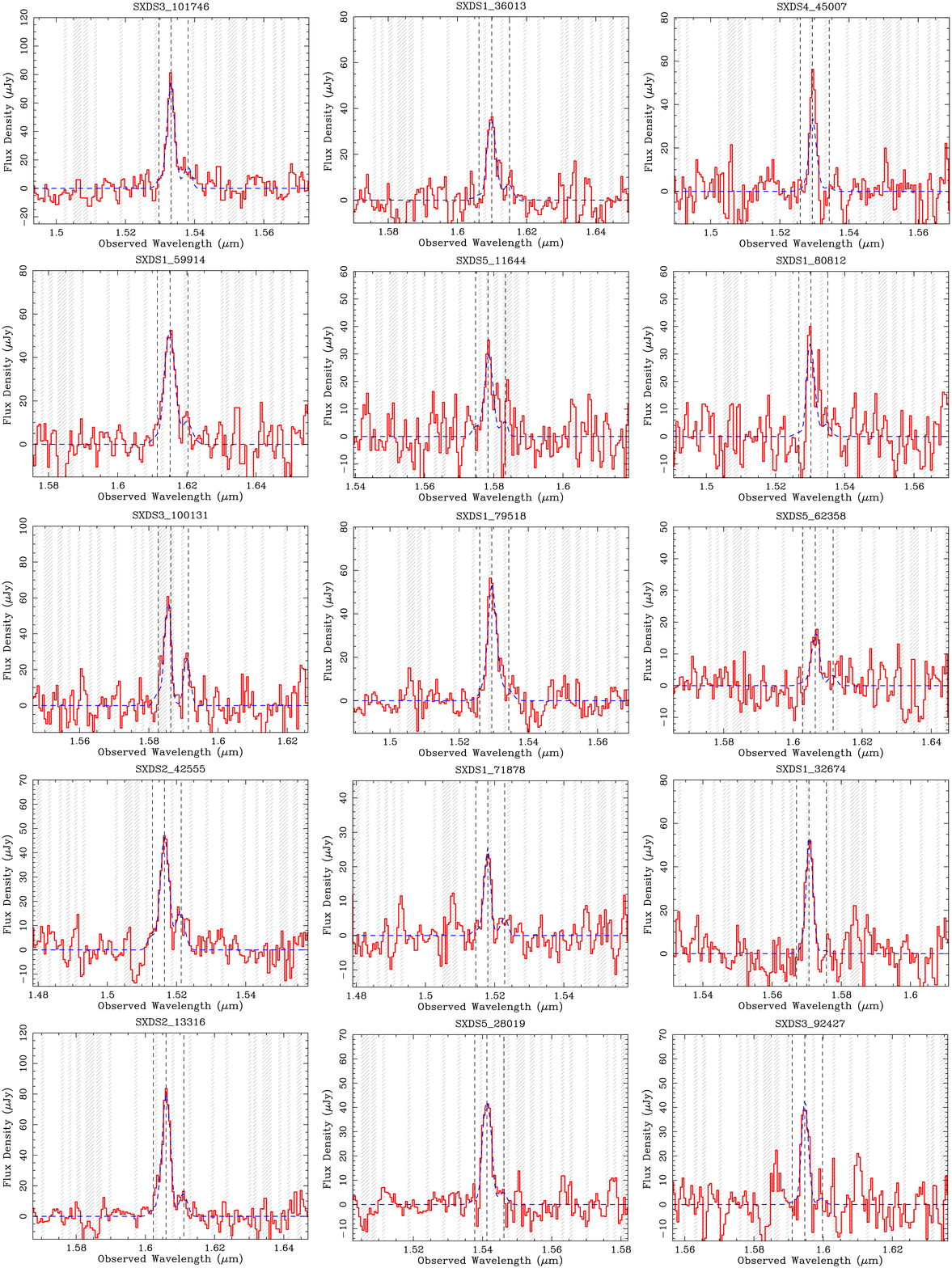}
\caption{Examples of the observed spectra (\textit{red solid lines}) with the best-fitted models (\textit{blue dotted lines}). In the left 5 panels, objects whose S/Ns of [N\emissiontype{II}]$\lambda$6584 lines are larger than 3 are plotted. In the middle panels, objects with 1.5 $\leq$ S/N $<$ 3.0, and in the right panels, objects with S/N $<$ 1.5 are plotted. Vertical \textit{dotted} lines indicate the positions of [N\emissiontype{II}]$\lambda$6548, H$\alpha$, and [N\emissiontype{II}]$\lambda$6584, from left to right. The positions and the width of the OH-mask are indicated by hatched stripes. Continua are subtracted by third-order polynomial fitting.\label{SpectraEx}}
\end{center}
\end{figure*}

In total, 317 objects were observed and for 71 objects the H$\alpha$ emission line was detected with a signal-to-noise ratio (S/N) larger than 3, where the noise level is measured from the continuum in a wavelength window of $\pm 0.1\mu$m from the emission line. Detection rates are 26\%, 44\%, 21\%, and 9\% at $M_{*} < 10^{10}M_{\odot}$, $10^{10}M_{\odot} \leq M_{*} < 10^{10.5}M_{\odot}$, $10^{10.5}M_{\odot} \leq M_{*} < 10^{11}M_{\odot}$, and $M_{*} \geq 10^{11}M_{\odot}$, respectively. Examples of the obtained spectra are presented in Figure \ref{SpectraEx}. {Spectra in which the S/N of [N\emissiontype{II}]$\lambda$6584 is larger than 3.0, 1.5 $\leq$ S/N $<$ 3.0, S/N $<$ 1.5 are shown in the left, middle, and right panels of the figure, respectively.}

\section{Results\label{Sec_Results}}
\subsection{Model Fitting to Emission Lines\label{Sec_FluxDet}}
In order to measure a spec-$z$, a flux of an emission line, and a line width, we made a model fit to the reduced one-dimensional emission line profile. This process is straightforward in the usual spectroscopy. In our case, however, we need rather complicated processes. FMOS uses an OH suppression system to remove the strong OH airglow sky emission lines. The dispersed light is focused onto the mask mirror with a resolving power of $R\sim3000$ ($\sim100$ km s$^{-1}$) and is rejected by the OH masks. Then the spectrum is anti-dispersed by VPH grating to $R\sim650$ ($\sim450$ km s$^{-1}$) and focused onto the detector (see \citet{Kimura:2010p11396} for the optical design of FMOS). Since the reduced spectra are divided by a spectrum of the calibration star in the relative flux calibration process of the pipeline, the effect of the OH-mask on the continuum light is corrected in the final reduced spectra. The correction, however, is not appropriate for emission lines if the line width is relatively small. Thus we present two different ways of the spectral fitting. One is fitting with model spectra taking into account the mask effects, and the other is fitting the observed spectra  without taking the mask effects into consideration. In both cases, we subtracted the continuum by fitting with third-order polynomials in the wavelength range of 1.45$-$1.70$\mu$m.

In the former method (\textit{method1}), we fit  model spectra including the OH-mask effect to the observed spectra. The model spectra are generated as follows: An intrinsic spectrum is modeled by a multi-Gaussian function with H$\alpha$ and [N\emissiontype{II}]$\lambda$$\lambda$6548,6584 lines. We assume that ratio of [N\emissiontype{II}]$\lambda$6584 and [N\emissiontype{II}]$\lambda$6548 is 3.0, the line widths are the same as that of H$\alpha$, and  there is no systematic shift between H$\alpha$ line and [N\emissiontype{II}] lines. Each spectrum is parametrized by four parameters: redshift, amplitude of H$\alpha$, amplitude of [N\emissiontype{II}]$\lambda$6584, and  line width. The spectrum represented by these parameters is convolved to a resolution of $R\sim3400$, and then, the flux density in the regions ($\Delta \lambda \sim 8\textrm{\AA}$) of the OH-masks is set to be zero. The obtained spectrum is convolved to a low-resolution spectrum with $R\sim800$.

\begin{figure}
\begin{center}
\includegraphics[scale=0.5,angle=270]{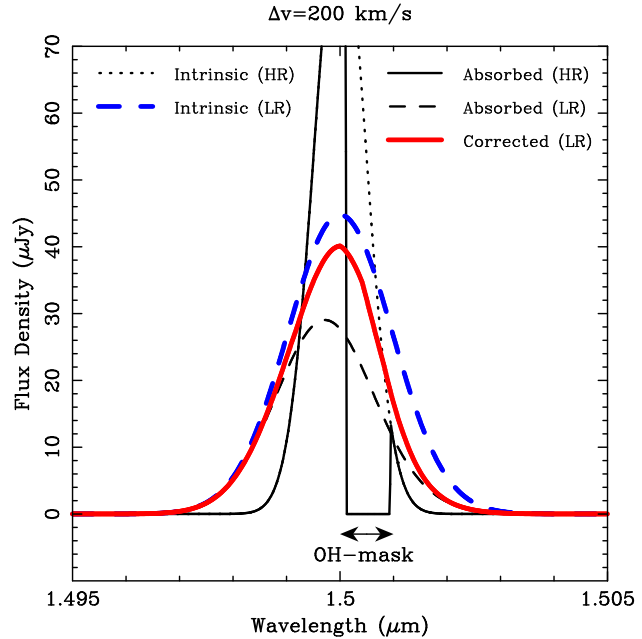}
\caption{Schematic view of the OH-mask effects. The intrinsic spectrum with $\Delta v$ of 200 km s$^{-1}$ in the HR mode and that absorbed by the OH-mask are indicated by a \textit{dotted line} and a \textit{thin solid line}, respectively. The absorbed spectrum degraded to the LR is indicated by a \textit{thin dashed line}. The spectrum corrected for the mask effects by using continuum light is indicated by a \textit{thick solid line}; this is the obtained spectrum after data reduction. The intrinsic spectrum degraded to the LR without considering the mask effect is shown by a \textit{thick dashed line} as a reference. See also text for details.\label{SpectraSim}}
\end{center}
\end{figure}

In order to correct the continuum level affected by the OH-suppression, a flat spectrum in $F_{\nu}$ without emission lines mimicking a star used in the flux calibration is processed in the same manner as  described above. The correction factors as a function of wavelength are derived from the resulting spectrum divided by the input flat spectrum. The model spectrum with emission lines is corrected by using this correction factor in order to compare with the observed spectrum. A schematic view of line profiles affected by the OH-mask is presented in Figure \ref{SpectraSim}.

The fitting process is carried out using \textit{mpfit} package \citep{Markwardt:2009p16616} which implements non-linear least square fitting with standard Levenberg-Marquardt algorithm. In the fitting procedure, we search for the minimum $\chi^{2}$ in the parameter space of the four parameters described above. The resulting $\chi^{2}$ maps of fits show that there are no pronounced discontinuities or strong degeneracies between the parameters.

For the latter method (\textit{method2}), we simply fit the observed spectrum by the multi-Gaussian profile. The fitting procedure is generally the same as \textit{method1}. The fitting models are multi-Gaussian functions, but without taking the mask loss corrections into consideration.

\begin{figure*}
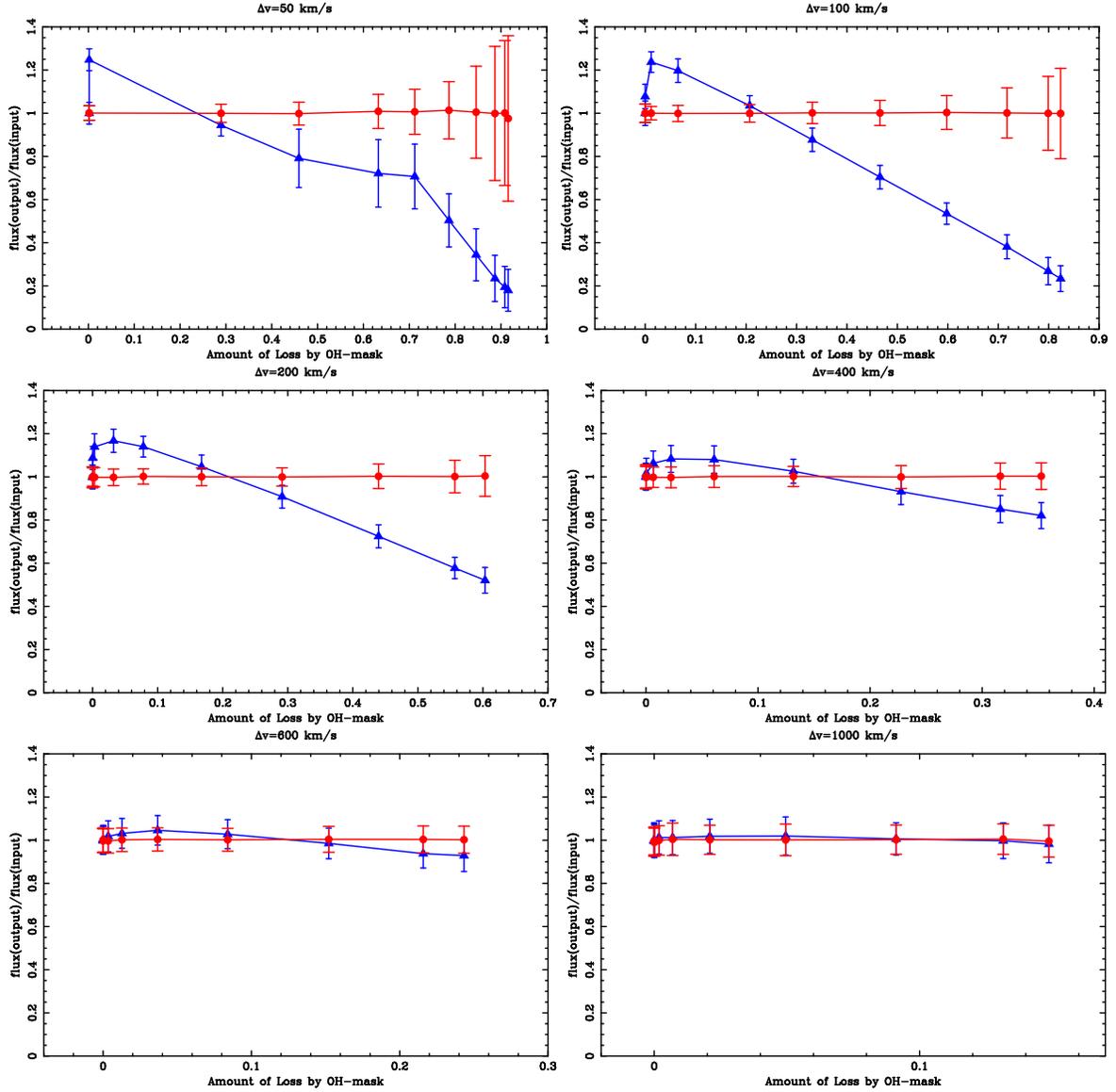

\begin{center}
\includegraphics[scale=0.5,angle=270]{f4a.eps}
\includegraphics[scale=0.5,angle=270]{f4b.eps}
\includegraphics[scale=0.5,angle=270]{f4c.eps}
\includegraphics[scale=0.5,angle=270]{f4d.eps}
\includegraphics[scale=0.5,angle=270]{f4e.eps}
\includegraphics[scale=0.5,angle=270]{f4f.eps}
\caption{The average ratio of recovered flux and the input flux against the amount of the loss by the OH-mask for intrinsic line widths of 50 km s$^{-1}$ (\textit{top left}), 100 km s$^{-1}$ (\textit{top right}), 200 km s$^{-1}$ (\textit{middle left}), 400 km s$^{-1}$ (\textit{middle right}), 600 km s$^{-1}$ (\textit{bottom left}), 1000 km s$^{-1}$ (\textit{bottom right}). \textit{Red lines} and \textit{circles} are the results by \textit{method1}, while the \textit{blue lines} and \textit{triangles} are those by \textit{method2}. Error bars indicate 1$\sigma$ standard deviations of the simulated distributions. See text for details of the simulations.\label{Fig_Sim}}
\end{center}
\end{figure*}

In order to quantify the effects of the OH-masks and to evaluate the accuracy of the two methods, we generate artificial spectra with one emission line and no continuum, varying the line widths (50 $-$ 1000 km s$^{-1}$) and fluxes (1.0$\times$10$^{-17}$ $-$ 1.0 $\times$ 10$^{-15}$ erg s$^{-1}$ cm$^{-2}$),  which cover the ranges of widths and fluxes of most of the emission lines detected with S/N $\geq$ 1.5. Then we place an OH-mask virtually changing its wavelength randomly to simulate the amount of flux loss for an emission line at different redshifts. The final LR spectrum is then made by degrading the spectral resolution down to $R\sim650$. The random noise, which is equivalent to the average noise level actually observed in a typical exposure time, is added into each spectrum. The resulting artificial spectra are fitted by two different methods (\textit{method1} and \textit{method2}), and the obtained redshifts (wavelengths), line fluxes, and line widths  are compared to the input values. We repeat this process 1000 times. The redshifts are recovered with an accuracy of $\lesssim5\%$ in both methods. The line widths are underestimated up to $\sim50\%$ in \textit{method2}. While, in \textit{method1}, the line widths can be recovered properly with an accuracy of $\lesssim20\%$.

The recovery rate of the flux against the amount of the loss by the mask is presented in Figure \ref{Fig_Sim}. Our simulations show that \textit{method2} cannot properly recover the flux which is affected by the OH-mask, especially when the line width is small. For a line width of $\sim$50 km s$^{-1}$, up to $\sim$90\% of the line flux is damped by the OH-mask. In this case, the flux recovered by \textit{method2} is only $\sim$20\% of the intrinsic flux, and $\sim80\%$ of the light is lost. The recovery of the flux improves if the line width is larger. For instance, the flux can be recovered within $\lesssim$20\% accuracy if the line width is $\gtrsim$600 km s$^{-1}$. It is also interesting to note that the flux is overestimated by \textit{method2} in some cases, because the line profile is overcorrected by the correction for the continuum: If a large part a pixel covers an OH-mask and the peak of a narrow emission line just outside of the mask, the flux loss rate in the pixel for the continuum is larger than that for the emission line.

Figure \ref{Fig_Sim} shows that fluxes recovered by \textit{method1} agree well with the intrinsic fluxes for any line width. The uncertainty of the correction is $\lesssim20\%$ in most cases. If the loss is $\gtrsim 90\%$, however, the uncertainty is $\gtrsim 30\%$. Of course, the results are obtained from the simulations and the actual uncertainty could be larger than these estimates.  We exclude objects from the observed sample, if the loss is calculated to be larger than 90\% in the profile fitting. In this paper, we use \textit{method1} as a fiducial line fitting method.

\subsection{Spec-$z$ and H$\alpha$ Line Flux\label{Sec_ParComp}}
\begin{figure*}
\begin{center}
\includegraphics[scale=0.75,angle=270]{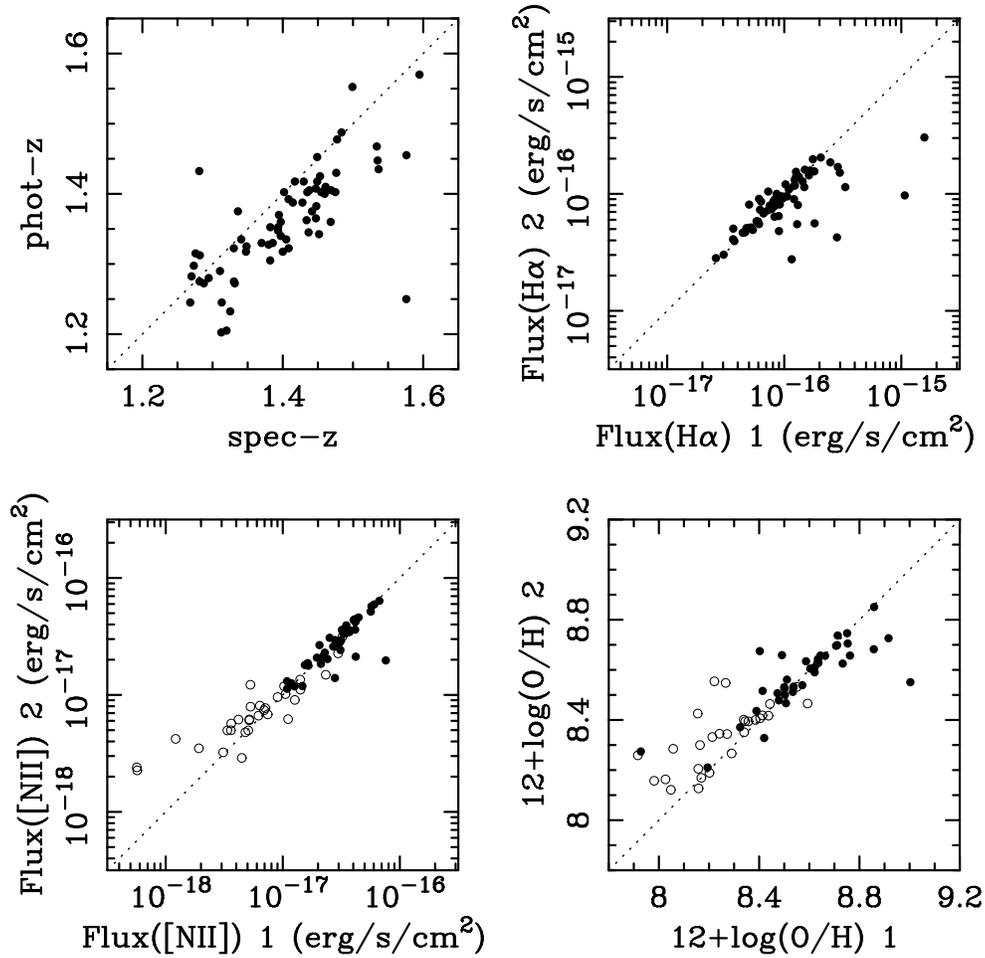}
\caption{Comparison between spec-$z$ and photo-$z$ (\textit{upper left}). Comparisons of H$\alpha$ flux (\textit{upper right}), [N\emissiontype{II}] flux (\textit{lower left}), and oxygen abundance (\textit{lower right}) in two different spectral fitting methods. Open circles are objects whose S/N of [N\emissiontype{II}] flux is less than 1.5.\label{ParComp}}
\end{center}
\end{figure*}

The best-fit redshifts, which are referred to as spec-$z$s, are distributed from $1.2\leq z \leq 1.6$ with median of $z=1.408$. In the upper left panel of Figure \ref{ParComp}, the spec-$z$ is compared to the photo-$z$. Although the spec-$z$ generally agrees with the photo-$z$ with $\sigma\sim0.05$, consistent with the uncertainty of the photo-$z$ determination described in Section \ref{Sec_Sample}, there appears to be a systematic difference of $\sim0.05$ at $1.2\leq z \leq 1.6$. This may be caused by the systematic uncertainty of the photometric zero-points. The presence of such systematic differences at some redshifts has been also reported by \citet{Ilbert:2006p454}.

In Figure \ref{ParComp}, we compare the fluxes of H$\alpha$ and [N\emissiontype{II}] emission lines obtained by the two different methods. Although the flux derived from the two methods generally agree well for both emission lines, in some cases the flux measured by \textit{method2} is underestimated significantly, which is consistent with the results of the simulations, indicating the necessity of the correction for the OH mask effects.

\begin{figure*}
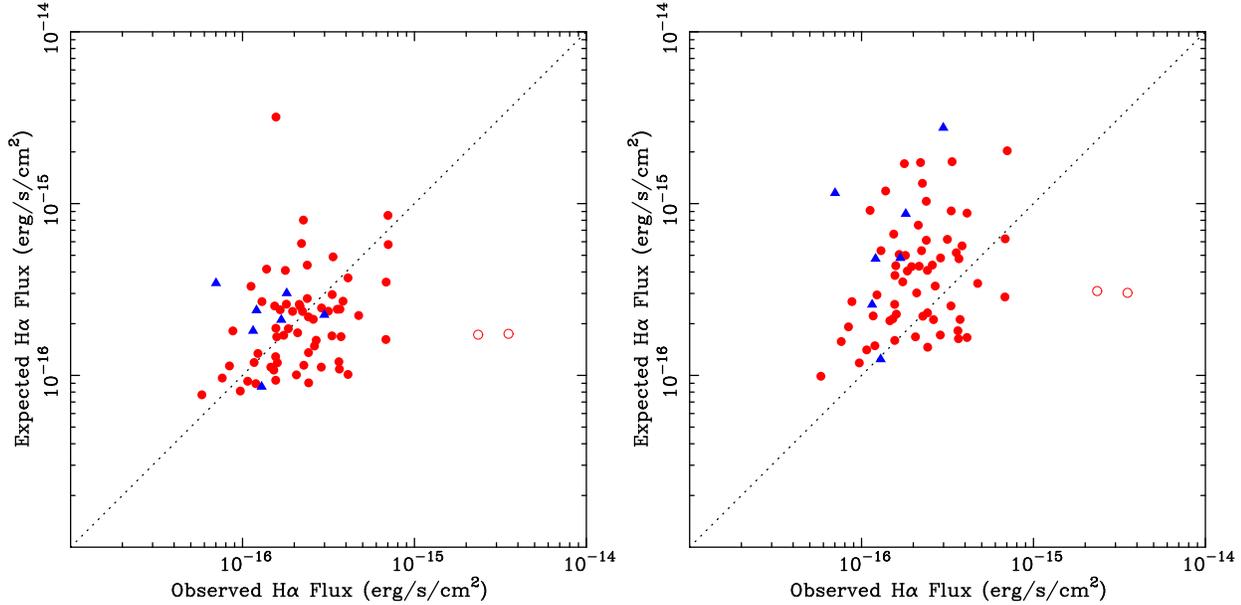

\begin{center}
\includegraphics[scale=0.45,angle=270]{f6a.eps}
\includegraphics[scale=0.45,angle=270]{f6b.eps}
\caption{Comparisons of the observed H$\alpha$ flux and the expected H$\alpha$ flux. The expected H$\alpha$ flux is calculated from the rest-frame UV luminosity corrected by $E(B-V)$ for nebular (\textit{left}) and stellar (\textit{right}) components assuming the \citet{Calzetti:2000p7012} extinction law. Objects with $>$90\% flux loss due to the OH-masks are indicated by \textit{red open circles}. The AGN candidates in the BPT diagrams are indicated by \textit{blue triangles} (Section \ref{Sec_AGN}).\label{FhaComp}}
\end{center}
\end{figure*}

\begin{figure}
\begin{center}
\includegraphics[scale=0.45,angle=270]{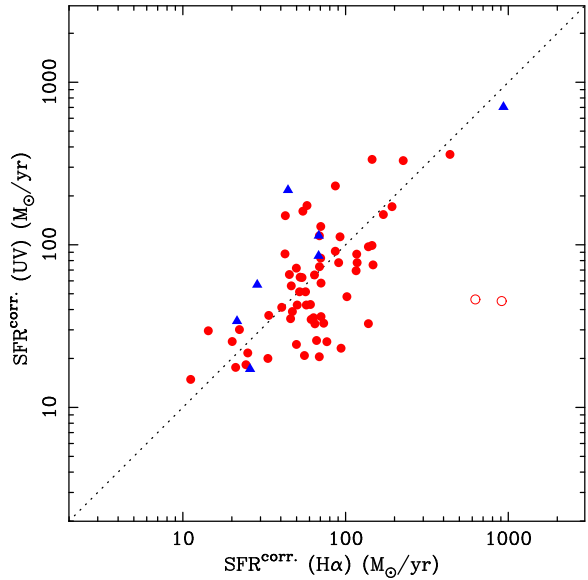}
\caption{Comparisons of the extinction corrected SFR obtained from the H$\alpha$ luminosity and that from the rest-frame UV luminosity density. Objects with $>$90\% flux loss due to the OH-masks are indicated by \textit{red open circles}. The AGN candidates in the BPT diagrams are indicated by \textit{blue triangles} (Section \ref{Sec_AGN}).\label{HaUVComp}}
\end{center}
\end{figure}

Comparisons of the observed H$\alpha$ flux and the expected H$\alpha$ flux are shown in Figure \ref{FhaComp}. The observed H$\alpha$ fluxes are corrected for the fiber loss depending on their $r_{50}$ as we describe in Section \ref{Sec_ApEff}. In the left panel, the expected H$\alpha$ fluxes calculated from the rest-frame UV luminosity, corrected for the nebular emission line using $E(B-V)$, agree well with the observed H$\alpha$ fluxes. In contrast, in the right panel, the expected fluxes by assuming the $E(B-V)$ for the stellar component are systematically larger than the observed ones, implying that the dust extinction for the nebular emission is larger that for the stellar continuum. Figure \ref{HaUVComp} shows that the SFR derived from the observed H$\alpha$ flux (extinction corrected for nebular component) agrees with that from the rest-frame UV luminosity density (extinction corrected for stellar component).

Since the H$\alpha$ emission line is superposed on the stellar absorption line, the correction for this effect is necessary in order to measure an accurate emission line flux. \citet{Zahid:2011p11939} fitted the stellar continuum of stacked spectra and found that the correction to H$\beta$ emission is negligible in galaxies at $z\sim0.8$. Since the continua of our sample are mostly too weak to be fitted, we estimate the absorption contribution from the SED models as follows. We calculate the equivalent width (EW) of H$\alpha$ absorption line by assuming an exponentially decaying  star formation history with $\tau=10$ Myr and a stellar age of $\sim100$ Myr; a typical case in our SED fitting for the observed sample. The resulting EW of $-4.2\textrm{\AA}$ is much smaller ($\sim2\%$) than the typical H$\alpha$ EW of $\sim200$\AA, and thus, we do not take the effect of stellar absorption into consideration.

\subsection{Aperture Effect\label{Sec_ApEff}}
A certain fraction of light outside the aperture of the FMOS fiber of \timeform{1.''2} diameter is lost. Although the aperture effect on relative quantities such as a line ratio and the metallicity is small, the effect is critical to absolute quantities such as total flux and SFR. We estimate the covering fraction for galaxies by using the NICMOS F160W archival images \citep{Bouwens:2010p17642} convolved to the typical seeing sizes(\timeform{0''.5}, \timeform{0''.75}, \timeform{1''.00}, and \timeform{1''.20} in FWHM) in the GOODS-N field. The object detection and \timeform{1''.2} aperture photometry are carried out by using \textit{SExtractor}. These objects are cross-matched with a K-selected catalogue in GOODS-N/MODS (\cite{Kajisawa:2010p7726,Yuma:2011p14330}). For objects brighter than $K_{s} =23.9$ mag and more massive than $M_{*}=10^{9.5} M_{\odot}$ at $1.2\leq z_{\rm ph} \leq 1.6$, we calculate the covering fraction, which is defined as FLUX\_APER (in \timeform{1''.2} diameter) / FLUX\_AUTO. The covering fraction correlates very well with a half light radius deconvolved by the PSF ($r_{50}$) in the F160W image; the scatter is very small ($\lesssim4\%$). According to the simulations, the covering fractions are $0.20-0.80$ depending on the seeing size; a typical value is $\sim0.5$ when the seeing size is \timeform{0''.75}.

In order to estimate fiber aperture losses for our sample galaxies in the SXDS field, we firstly derive the observed $r_{50}$ based on the K-band WFCAM images by using the SExtractor's FLUX\_RADIUS. The intrinsic $r_{50}$ of our sample is calculated from the observed $r_{50}$ from the K-band image, after deconvolution by a Gaussian PSF of \timeform{0''.80}. The resulting $r_{50}$ distributes from \timeform{0''.3} to \timeform{0''.9} with the median value of \timeform{0''.55} (4.6 kpc at $z=1.4$). In order to examine the reliability of the estimation in K-band, these are compared to those obtained from higher-resolution images from WFC3/HST in the CANDELS survey held in a small part of the SXDS/UDS field. \citep{Grogin:2011p17008, Koekemoer:2011p17029}. The $r_{50}$ of our primary sample, described in Section \ref{Sec_Selection}, in the CANDELS FoV is  measured from the F160W (the rest-frame wavelength is slightly different from that in K-band) images deconvolved by a PSF of \timeform{0''.18}. We found that they agree well with those derived from the K-band image within \timeform{0''.05} ($\sim 10 \%$ accuracy).

The fiber loss for each object is determined from its $r_{50}$. The seeing size is assumed to be \timeform{0''.75} (FWHM) in $H$-band, as expected for the seeing size of $\sim$\timeform{0''.9} in $R$-band and the wavelength dependence $\propto \lambda^{-0.2}$ (e.g., \cite{Glass:1999}). The median $r_{50}$ is \timeform{0''.55} and the corresponding covering fraction is 0.45. Line fluxes that we present in this paper are corrected for this aperture effect\footnote{Strictly speaking, the aperture loss is already corrected by using the correction factor for a point source in the flux calibration. The correction factor should be different for a galaxy since the galaxy is generally extended. Although we need to modify the correction factor for a galaxy, the modification is less than $10\%$ and we did not modify it.}. It should be noted that the positional error of the fibre allocation is $\lesssim$\timeform{0''.2} \citep{Kimura:2010p11396} and the effect is small ($\lesssim$10\%). Further, this error is random and does not cause a systematic error.

\subsection{Possible AGNs \label{Sec_AGN}}
\begin{figure*}
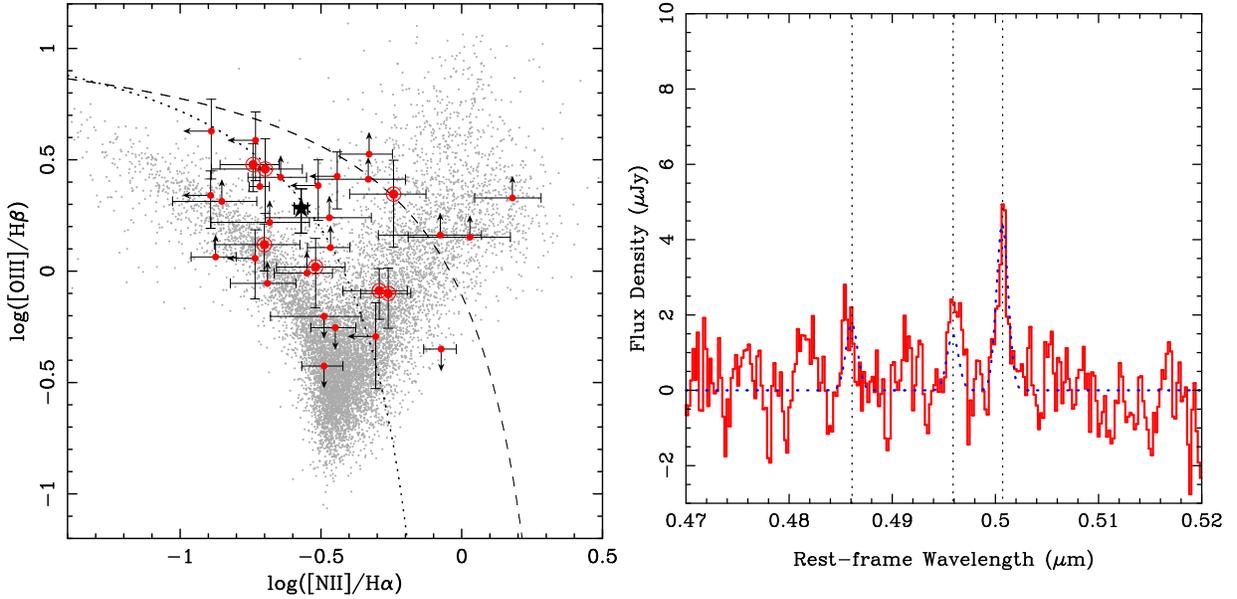

\begin{center}
\includegraphics[scale=0.5,angle=270]{f8a.eps}
\includegraphics[scale=0.45,angle=270]{f8b.eps}
\caption{\textit{(left)}: BPT diagram. Upper and lower limits are 3$\sigma$ values. The result from the stacking analysis is shown by a black star. For the comparison, local SDSS galaxies are plotted as gray dots. The empirical criterion to separate the AGN and SF by \citet{Kauffmann:2003p8323} and the maximum theoretical line of starbursts by \citet{Kewley:2001p5158} are shown as dotted and dashed curves, respectively. \textit{(right)}: The stacked non-AGN spectrum for the H$\beta$ and [O\emissiontype{III}] region.\label{BPT}}
\end{center}
\end{figure*}

Detections of H$\alpha$, [N\emissiontype{II}]$\lambda6584$, [O\emissiontype{III}]$\lambda5007$, and H$\beta$ lines allow us to make a line diagnosis to distinguish AGNs from star-forming (SF) galaxies \citep{Baldwin:1981p15095}. H$\alpha$, [N\emissiontype{II}], [O\emissiontype{III}], and H$\beta$ lines are detected with S/N $>$ 3 for 7 objects. H$\alpha$ and [N\emissiontype{II}] lines are detected with S/N $>$ 3 but both [O\emissiontype{III}] and H$\beta$ lines are not detected for 14 objects. Both [O\emissiontype{III}] and H$\beta$ lines are detected with S/N $>$ 3 but H$\alpha$ and [N\emissiontype{II}] lines are not detected for 4 objects. The distribution of these objects on the BPT diagram is presented in Figure \ref{BPT}. The empirical criterion to separate the AGN and SF by \citet{Kauffmann:2003p8323} and the maximum theoretical line of starburst by \citet{Kewley:2001p5158} are plotted. Six objects are in the AGN region if the theoretical line is adopted. In addition to these objects, one object shows a line width larger than 1000 km s$^{-1}$, suggesting the presence of an AGN. Thus these seven objects are excluded from the sample hereafter. It is worth stressing that although several objects are located between the theoretical line and the empirical line, often referred to as a \textit{composite} region, most are within the empirical line. 

In the right panel of Figure \ref{BPT}, we also show the stacked spectrum of all observed spectra excluding  the seven AGN candidates. The stacking analysis shows that these galaxies at $z\sim1.4$ are located above the sequence of normal SF galaxies at $z\sim0.1$.  Similar trends have been reported at $z\sim2$ \citep{Erb:2006p4143, Hainline:2009p7444}. By using high S/N spectra of gravitationally lensed galaxies, \citet{Hainline:2009p7444} found that the galaxies at $z\sim2$ show higher [O\emissiontype{III}]/[O\emissiontype{II}] line ratios and thus higher ionization parameters compared to local SF galaxies. The systematic upward shift of our sample in the BPT diagram also suggests higher ionization parameters in galaxies at $z\sim1.4$. For this reason, we use the maximum theoretical line by \citet{Kewley:2001p5158} as the selection of non-AGNs. However, we cannot completely rule out the possibility of contamination from low-luminosity AGNs.

\subsection{Metallicity Determination and Its Uncertainty}
Empirical indicators obtained from strong emission lines are used in order to derive the metallicity of high-redshift galaxies. The choice of the indicators depends on the targets' redshift. For galaxies at $1.2<z<1.6$, H$\beta$, [O\emissiontype{III}], H$\alpha$, and [N\emissiontype{II}] can be covered  in the FMOS wavelength range.

A metallicity indicator with the line ratio of [N\emissiontype{II}] and H$\alpha$ (N2 method) was calibrated by the following equation \citep{Pettini:2004p7356}:
\begin{equation}
12+\textrm{log(O/H)}=8.90+0.57 \times \textrm{N2},
\end{equation}
where N2 refers to log[$f(\textrm{[N\emissiontype{II}]}\lambda6584)$/$f(\textrm{H}\alpha$)]. The scatter around the relation is $\sim0.18$ dex at 68\% significance level \citep{Pettini:2004p7356}. It is also known that [N\emissiontype{II}]  tends to be saturated near and above solar metallicity. \citet{Pettini:2004p7356} found that O3N2 ($\equiv$ log\{([O\emissiontype{III}]$\lambda$5007/H$\beta$)/([N\emissiontype{II}]$\lambda$6584/H$\alpha$)\}) is more useful for higher metallicities. For 14 objects in our sample, the [N\emissiontype{II}], H$\alpha$, [O\emissiontype{III}] and H$\beta$ lines are detected with S/N $\geq$ 2. For these objects, we compared the metallicity derived by the N2 method to that from the O3N2 method. We found that the metallicities derived from the O3N2 method are systematically lower by $0.07$ dex than those from N2 method in the metallicity range of $12+\textrm{log(O/H)}=8.4-8.8$. The systematic difference of metallicity determination in these two methods by $0.1-0.2$ dex was  reported at high-redshift \citep{Erb:2006p4143}; this may be due to differing physical conditions in SF galaxies at $z\sim 1.4$ as compared to $z\sim0$.

In Figure \ref{ParComp}, the metallicity derived from \textit{method1} is compared to that from \textit{method2}. Although the metallicities derived from the two methods generally agree, the scatter is relatively large.

\subsection{Stacking Analysis \label{Sec_Stack}}
More than half of the [N\emissiontype{II}] line fluxes are weak to measure the [N\emissiontype{II}]/H$\alpha$ line ratio and thus metallicity of individual objects reliably. In order to cope with this low significance of the [N\emissiontype{II}] line and to reveal the 'average' nature of the sample galaxies, we separate the sample into three mass bins, stack the individual spectra, and derive the metallicity at the representative stellar mass in each bin. We perform the stacking analysis in two ways. One is stacking the best-fit spectra which are properly corrected for the OH-mask effect as described above, and the other is simply stacking the observed spectra.

For the former method (stacking the model spectra), each model spectrum is made by a multi-Gaussian profile with the best-fit parameters from \textit{method1} (Section \ref{Sec_FluxDet}). These spectra are combined to give the average spectrum. The obtained spectra are again fitted by multi-Gaussian models  to derive the line ratio and the metallicity.

\begin{figure*}
\begin{center}
\includegraphics[scale=0.8,angle=270]{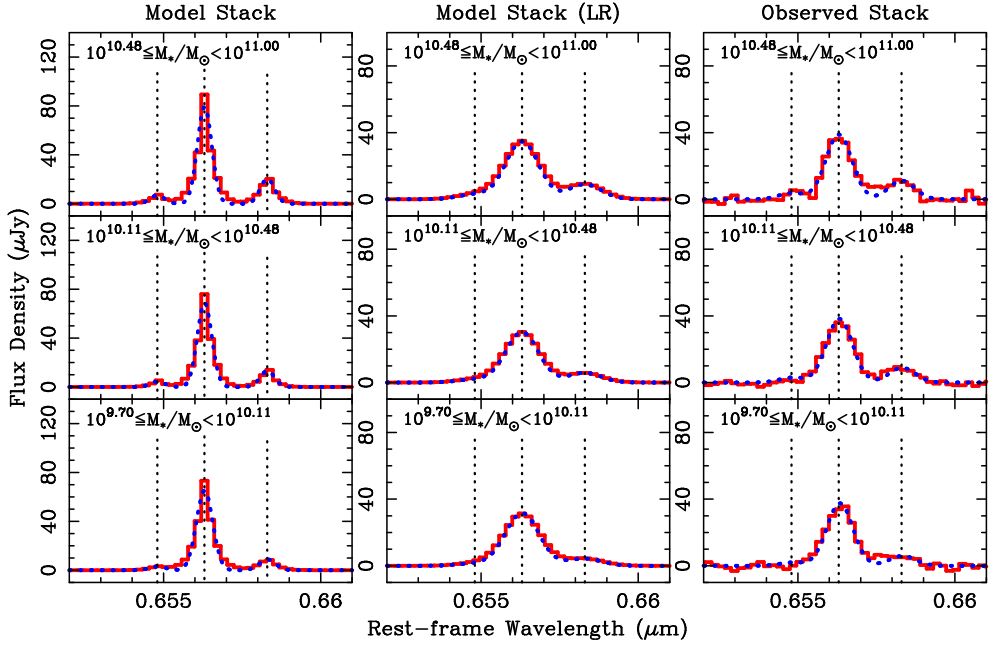}
\caption{Stacked spectra in three mass bins (\textit{red solid line}) and the model fittings (\textit{blue dotted line}). The profiles from the stacking analysis with best-fit models are in the left panels, those with best-fit models convolved to the LR spectra in the middle, and those with the ``observed spectra'' in the right panel. The mass ranges (the number of objects in each mass bin) are 10$^{10.48}$ $<$ $M_{*}$/$M_{\odot}$ $<$ 10$^{11.00}$ (14), 10$^{10.11}$ $<$ $M_{*}$/$M_{\odot}$ $<$ 10$^{10.48}$ (22), and 10$^{9.70}$ $<$ $M_{*}$/$M_{\odot}$ $<$ 10$^{10.11}$ (22) from top to bottom. Vertical \textit{dotted} lines indicate the positions of [N\emissiontype{II}]$\lambda$6548, H$\alpha$, and [N\emissiontype{II}]$\lambda$6584 from left to right.\label{stack1}}
\end{center}
\end{figure*}

For the latter method (stacking the observed spectra), the observed spectra  are averaged with weights based on the observed noise to minimize the effects of the OH-mask as follows.

\begin{equation}
F_{2}^{stack}(\lambda)=\sum_{i=1}^{n} \frac{F_{i}(\lambda)}{\sigma_{i}(\lambda)^{2}}/\sum_{i=1}^{n}\frac{1}{\sigma_{i}(\lambda)^{2}},
\end{equation}
where $F_{i}(\lambda)$ and $\sigma_{i}(\lambda)$ are the de-redshifted observed flux density and noise, respectively. The noise is generally large at the OH masks. In the left panel of Figure \ref{stack1}, the spectra resulting from stacking the model spectra are shown. In the middle panel, the spectra are convolved to low resolution ($R\sim800$ at the wavelength). In the right panel, the spectra resulting from stacking the observed spectra are shown. In the low resolution spectra, the spectra from the two different methods are roughly comparable. In Figure \ref{stack1}, it can be seen that the flux of [N\emissiontype{II}] increases with the increasing stellar mass.

The uncertainties associated with the stacking methods are examined  by stacking the sample in various ways. For both methods, we examine four different ways of combining spectra: The weighted average, unweighted average, median, and flux normalized average. The typical uncertainties are 0.04 dex (the largest mass bin) to 0.07 dex (the smallest mass bin). The uncertainties from the stacking method are almost comparable to or slightly larger than the observational errors. 

In the method of stacking observed spectra, some of the individual emission lines are affected by the OH-masks.This effect is simulated by stacking artificial spectra whose redshift is randomly distributed in the range of $1.2<z<1.6$. In this simulation, the line flux of H$\alpha$, the ratio of [N\emissiontype{II}]/H$\alpha$, and the line width are assumed to be $1.0 \times 10^{-16}$ erg s$^{-1}$ cm$^{-2}$, 0.20, and 200 km s$^{-1}$, respectively, which are typical values for the detected objects. The number of the artificial spectra is $\sim20$, which is comparable to the number of the actual objects in each mass bin. We fit the realized spectrum and measure the line ratio and metallicity by \textit{method2} described in Section \ref{Sec_FluxDet}. This process is repeated 1000 times. The recovered line ratio agrees with the input value within $\sim20\%$, corresponding to $\sim0.1$ dex in metallicity, with no systematic difference. 

In this paper, we use the former method, i.e., stacking the model spectra, as a fiducial stacking method. Hereafter, the stacked results are obtained by this method unless otherwise noted. 

\begin{figure*}
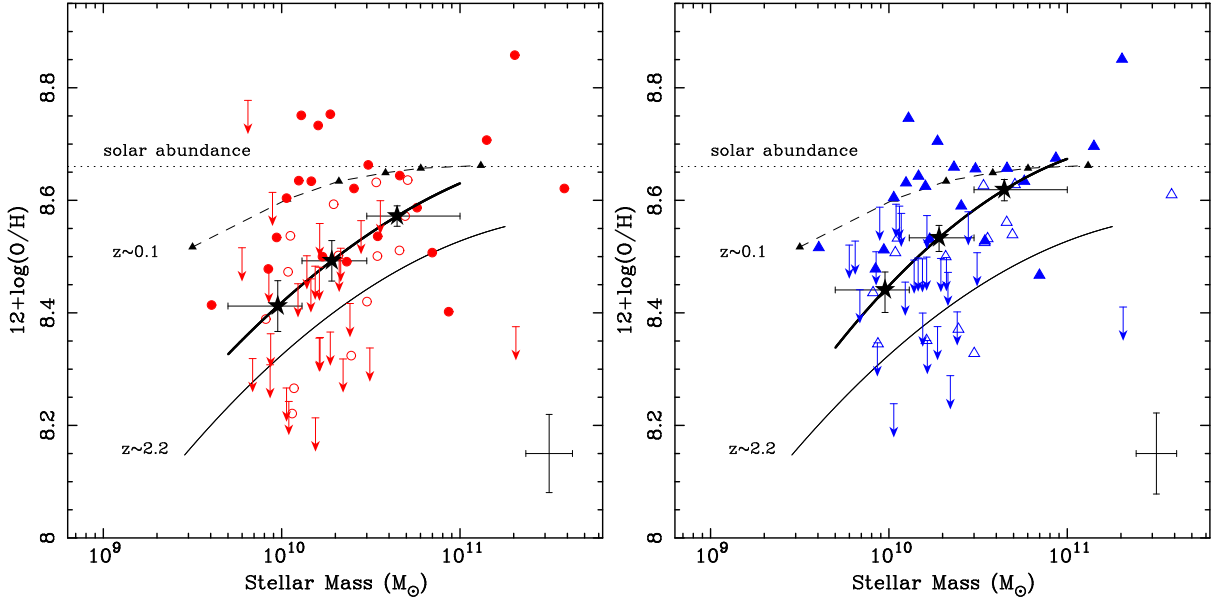

\begin{center}
\includegraphics[scale=0.5,angle=270]{f10a.eps}
\includegraphics[scale=0.5,angle=270]{f10b.eps}
\caption{Metallicity as a function of stellar mass of our sample. The metallicity derived with spectral fitting with and without the effects of the OH masks are plotted in the \textit{left} and \textit{right} panel, respectively. The typical errors of stellar mass and metallicity are shown in the lower right corner of each panel. [N\emissiontype{II}] lines with S/N $\geq$ 3.0 and 1.5 $\leq$ S/N $<$ 3.0 are indicated by \textit{filled} and \textit{open} symbols, respectively. [N\emissiontype{II}] lines with S/N $<$ 1.5 are plotted as upper limits with values corresponding to $1.5\sigma$. Note that the error of the metallicity is derived from the flux error of H$\alpha$ and [N\emissiontype{II}] lines and does not include the uncertainty of the metallicity calibration. The results from stacking analysis are presented by \textit{filled stars}. \textit{Thick solid lines} are the second-order polynomials for the stacked results. \textit{Dashed lines} and \textit{thin solid lines} are the mass-metallicity relation at $z\sim0.1$ \citep{Tremonti:2004p4119} and $z\sim2.2$ \citep{Erb:2006p4143}. Horizontal dotted line indicates solar metallicity (12+log(O/H)=8.66; \cite{Asplund:2004p11879}).\label{MZR}}
\end{center}
\end{figure*}

\begin{table}
 \begin{center}
 \caption{Metallicity from the stacked spectra in three stellar mass bins}\label{Tab1}
 \begin{tabular}{ccc}
 \hline \hline
log($M_{*}$/$M_{\odot}$) & \multicolumn{2}{c}{12+log(O/H)}\\
\cline{2-3}
& method1 & method2 \\
\hline
9.98$_{-0.28}^{+0.13}$ & 8.412$\pm$0.045 & 8.441$_{-0.037}^{+0.032}$\\
10.28$_{-0.17}^{+0.20}$ & 8.492$\pm$0.036 & 8.530$_{-0.024}^{+0.022}$\\
10.65$_{-0.17}^{+0.35}$ & 8.572$\pm$0.018 & 8.619$_{-0.020}^{+0.018}$\\
\hline
\end{tabular}
\end{center}
\end{table}

\subsection{The Mass-Metallicity Relation at $z\sim1.4$\label{Sec_MZR}}
The derived metallicity distribution as a function of stellar mass is presented in Figure \ref{MZR}. The stellar masses are re-calculated by the SED fitting in the same manner described in Section \ref{Sec_Sample} but fixing the redshift to the spectroscopic redshift obtained by the observations. All other physical quantities are based on the spec-$z$. In the left panel, metallicities are derived using \textit{method1} as we described in Section \ref{Sec_Stack}, while in the right panel, we use metallicities derived by \textit{method2}. In both panels, the objects for which the S/N of the [N\emissiontype{II}] flux is less than 1.5 are plotted as upper limits with values corresponding to $1.5\sigma$. The mass-metallicity relations obtained at $z\sim0.1$ \citep{Tremonti:2004p4119} and $z\sim2.2$ \citep{Erb:2006p4143} are plotted for comparison. Note that the stellar masses from these works are converted by a factor of 1.8 so that the IMF is consistent with that adopted in this paper. The metallicity at $z\sim2.2$ is obtained by using the same N2 indicator, and for that at $z\sim0.1$ we use the recalculated values with the N2 indicator from \citet{Erb:2006p4143} for a fair comparison. In both panels of Figure \ref{MZR}, typical observational errors of stellar mass and metallicity are shown at the lower-right corner. In these errors, the uncertainty of the metallicity calibration is not included.

Since  [N\emissiontype{II}] emission lines are not detected significantly for $\sim40\%$ of our sample, we measure the average metallicity by dividing the sample by stellar mass and applying stacking analysis including both detections and non-detections in the way presented in Section \ref{Sec_Stack}. We divide the sample into three mass bins, $10^{9.70}M_{\odot}-10^{10.11}M_{\odot}$, $10^{10.11}M_{\odot}-10^{10.48}M_{\odot}$, and $10^{10.48}M_{\odot}-10^{11.00}M_{\odot}$. The number of objects in each mass bin is $\sim$20. The metallicities derived from stacked spectra in three stellar mass bins are also presented as solid stars in the both panels of Figure \ref{MZR}. The errors are estimated by the bootstrap method with 1000 realizations. The resulting mass-metallicity relation obtained by our fiducial method (left panel) is systematically lower by 0.03, 0.04, and 0.05 dex, in the lowest, middle, and highest mass bin, respectively, than that from the observed spectra stacking method. The results are also summarized in Table \ref{Tab1}.

Figure \ref{MZR} shows that most of individual objects and the results from the stacking analysis at $z\sim1.4$ lie between the mass-metallicity relation at $z\sim0.1$ and  $z\sim2.2$. The stacked results are expressed by a second-order polynomial:
\begin{equation}
12+\textrm{log(O/H)}=-0.0724x^{2}+1.733x-1.67 \label{Eq_MZR1},
\end{equation}
for the model spectra stacking method and
\begin{equation}
12+\textrm{log(O/H)}=-0.1082x^{2}+2.497x-5.71 \label{Eq_MZR2},
\end{equation}
for the observed spectra stacking method, where $x=\textrm{log}(M_{*}/M_{\odot})$ in both equations. The resulting curves are also presented in both panels of Figure \ref{MZR}.

In section \ref{Sec_ApEff}, we found that the typical covering fraction of FMOS fiber is $\sim0.5$. The covering fraction by the fibre aperture affects the estimation of metallicity if there is a steep radial metallicity gradient; a small covering fraction leads to an overestimation of the metallicity if the metallicity gradient is negative. \citet{Tremonti:2004p4119} showed that the aperture effects on the metallicity is $\sim0.1$ dex by using the SDSS galaxies with the typical covering fraction of 0.24. In the local universe, the difference between the metallicity measured in the aperture and the integrated metallicity is $\lesssim0.1$ dex for the covering fraction of $\gtrsim0.2$ \citep{Kewley:2005p5336}. Since the typical covering fraction of our sample is $\sim0.5$, the metallicity gradient is unlikely to affect the derived metallicity so much unless the metallicity gradient drastically changes at high-redshift. Recently, steeper metallicity gradients than local galaxies have been reported at $z\sim 2$ \citep{Jones:2010p17052,Yuan:2011p9570}, while the flatter or even ``positive'' gradients have also been reported \citep{Cresci:2010p17086}. The metallicity gradient at high redshift remains unclear due to the small size of currently available samples.

\begin{figure}
\begin{center}
\includegraphics[scale=0.5,angle=270]{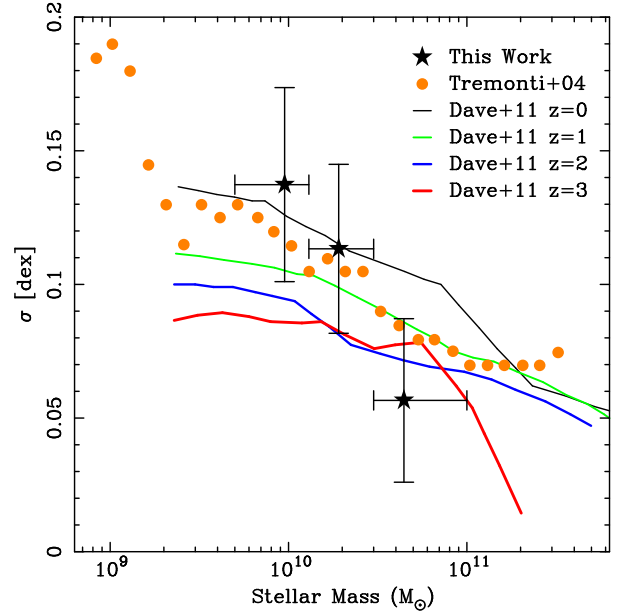}
\caption{Intrinsic scatter of the mass-metallicity relation at $z\sim1.4$ (\textit{filled stars}). Strictly speaking, this is a lower limit of the scatter. Error bars are based on the bootstrap method. Scatters at $z\sim0.1$ by \citet{Tremonti:2004p4119} are indicated by \textit{filled circles} and theoretical predictions at $z=0$, 1, 2, and 3 by \citet{Dave:2011p11823} are indicated by \textit{solid lines}.\label{MZR_Scatter}}
\end{center}
\end{figure}

\begin{figure*}
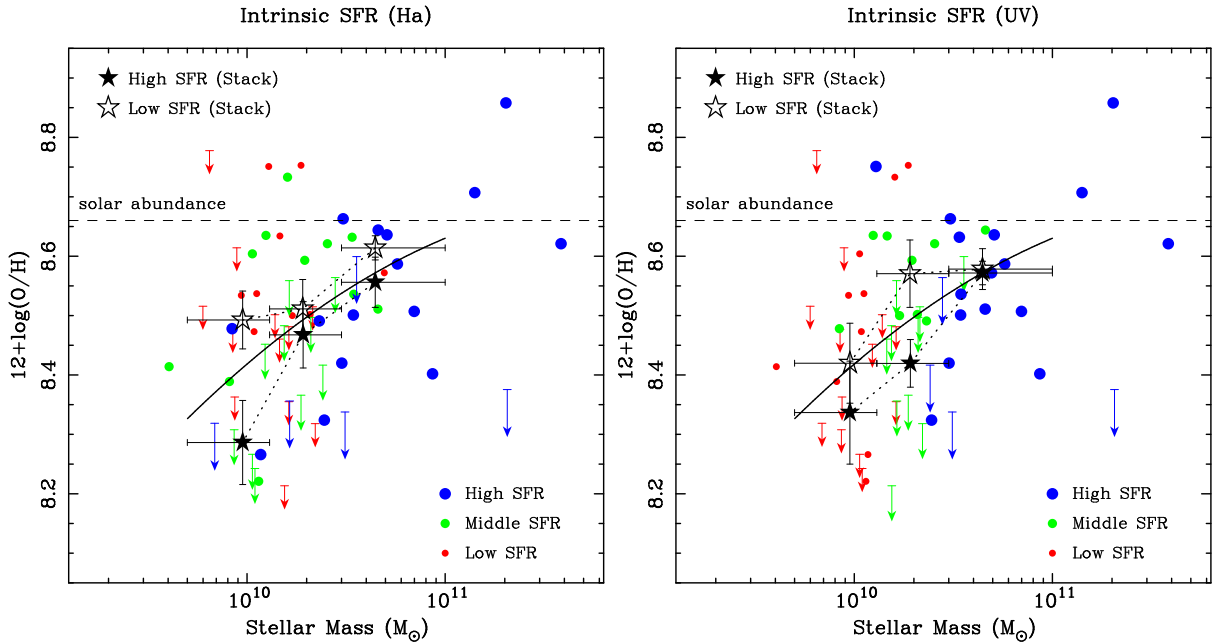

\begin{center}
\includegraphics[scale=0.5,angle=270]{f12a.eps}
\includegraphics[scale=0.5,angle=270]{f12b.eps}
\caption{Dependence of SFR on the mass-metallicity relation. In the left panel, the sample is divided into three groups with their SFRs derived from H$\alpha$ luminosity corrected for dust extinction: \textit{Circles} for detection and \textit{arrows} for non-detection for [N\emissiontype{II}] with SFR $\leq$ 53.0 $M_{\odot}$yr$^{-1}$ (\textit{small red}), 53.0 $<$ SFR $\leq$ 85.0 $M_{\odot}$yr$^{-1}$ (\textit{middle green}), SFR $>$ 85.0 $M_{\odot}$yr$^{-1}$ (\textit{large blue}). In the right panel, the sample is divided by their SFRs derived from the rest-frame UV luminosity density corrected for dust extinction, with SFR $\leq$ 40.0 $M_{\odot}$yr$^{-1}$ (\textit{small red}), 40.0 $<$ SFR $\leq$ 80.0 $M_{\odot}$yr$^{-1}$ (\textit{middle green}), SFR $>$ 80.0 $M_{\odot}$yr$^{-1}$ (\textit{large blue}). The \textit{filled stars} show the stacked results in the higher SFR group, while the \textit{open stars} show those in the lower SFR group in each mass bin. \textit{Solid curves} show the mass-metallicity relations described by Eq. \ref{Eq_MZR1} in Section \ref{Sec_MZR}.\label{MZR_IntSFR}}
\end{center}
\end{figure*}

The intrinsic scatter of the mass-metallicity relation has been reported at $z\sim0.1$ \citep{Tremonti:2004p4119, Yates:2011p16030} and also reproduced by cosmological simulations \citep{Dave:2011p11823}. We try to constrain the scatter of the mass-metallicity relation at $z\sim 1.4$, though the sample size may not be large enough yet. Figure \ref{MZR} shows that the metallicity distribution has a larger scatter as compared with the typical observational error.  So the scatter is defined as the RMS of the deviation from the best-fit mass-metallicity relation (Eq. \ref{Eq_MZR1}) in each mass bin, after subtracting the contribution from the averaged observational error of individual objects. The resulting standard deviations are $\sigma=$ 0.138, 0.113, and 0.057 dex in mass bins of log($M_{*}$/$M_{\odot}$) $=$ 9.98, 10.28, 10.65, respectively. Since we take the metallicities from upper limits for many objects, the intrinsic scatter should be larger than these values. Figure \ref{MZR_Scatter} shows the intrinsic scatter against stellar mass. The scatters of the mass-metallicity relation at $z\sim1.4$ are comparable to those obtained at $z\sim0.1$ by \citet{Tremonti:2004p4119}; they decrease with increasing stellar mass of galaxies if we see the face values, though the small scatter at the massive part may be partly due to the saturation effect of the N2 indicator.

\begin{figure*}
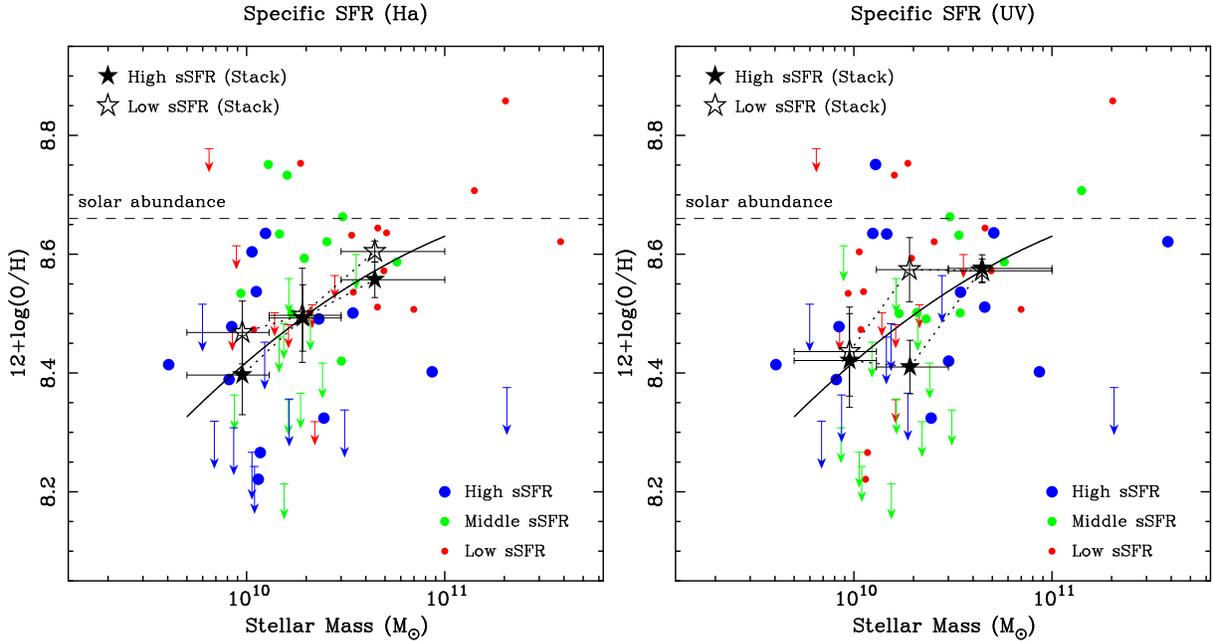

\begin{center}
\includegraphics[scale=0.5,angle=270]{f13a.eps}
\includegraphics[scale=0.5,angle=270]{f13b.eps}
\caption{Dependence of specific SFR (sSFR) on the mass-metallicity relation. In the left panel, the sample is divided into three groups according to sSFRs derived from H$\alpha$ luminosity corrected for dust extinction:  \textit{Circles} for detection and \textit{arrows} for non-detection for [N\emissiontype{II}] with sSFR $\leq$ 2.5 Gyr$^{-1}$ (\textit{small red}), 2.5 Gyr$^{-1}$ $<$ sSFR $\leq$ 4.0 Gyr$^{-1}$ (\textit{middle green}), sSFR $>$ 4.0 Gyr$^{-1}$ (\textit{large blue}). In the right panel, the sample is divided according to sSFRs derived from the rest-frame UV luminosity density corrected for dust extinction, with sSFR $\leq$ 2.3 Gyr$^{-1}$ (\textit{small red}), 2.3 Gyr$^{-1}$ $<$ sSFR $\leq$ 3.6 Gyr$^{-1}$ (\textit{middle green}), sSFR $>$ 3.6 Gyr$^{-1}$ (\textit{large blue}). The \textit{filled stars} show the stacked results in the larger sSFR groups, while the \textit{open stars} show those in the smaller sSFR groups in each mass bin. \textit{Solid curves} show the mass-metallicity relations described by Eq. \ref{Eq_MZR1} in Section \ref{Sec_MZR}.\label{MZR_sSFR}}
\end{center}
\end{figure*}

\begin{table*}
 \begin{center}
 \caption{Metallicity from the stacked spectra in three stellar mass bins and two SFR groups}\label{Tab2}
 \begin{tabular}{ccccccc}
 \hline \hline
log($M_{*}$/$M_{\odot}$) & SFR$^{\textrm{med}}$ (H$\alpha$) & SFR$^{\textrm{med}}$ (UV) & \multicolumn{4}{c}{12+log(O/H)}\\
\cline{4-7}
& ($M_{\odot}$yr$^{-1}$) & ($M_{\odot}$yr$^{-1}$) & Low SFR (H$\alpha$) & High SFR (H$\alpha$) & Low SFR (UV) & High SFR (UV)\\
\hline
9.98$_{-0.28}^{+0.13}$   & 56.7 & 25.8 & 8.493$\pm$0.049 & 8.286$\pm$0.071 & 8.420$\pm$0.067 & 8.337$\pm$0.087 \\
10.28$_{-0.17}^{+0.20}$ & 55.9 & 54.0 & 8.511$\pm$0.050 & 8.468$\pm$0.056 & 8.571$\pm$0.057 & 8.420$\pm$0.040 \\
10.65$_{-0.17}^{+0.35}$ & 112.9 & 91.5 & 8.614$\pm$0.020 & 8.556$\pm$0.042 & 8.578$\pm$0.034 & 8.572$\pm$0.020 \\
\hline
\end{tabular}
\end{center}
\end{table*}

\subsection{Dependence of SFR and Size on the Mass-Metallicity Relation \label{Sec_Dependence}}

We examine the dependency of SFR on the mass-metallicity relation for our sample at $z\sim1.4$. In Figure \ref{MZR_IntSFR}, we present the metallicity as a function of stellar mass for individual objects dividing into three groups according to their intrinsic SFRs. In the left panel, the SFR is derived from the observed H$\alpha$ luminosity corrected for extinction by using $E(B-V)^{gas}$ described in Section \ref{Sec_Sample}, while in the right panel, the SFR is derived from the rest-frame UV luminosity density also corrected for extinction. They show no clear trend that galaxies with higher SFRs show lower metallicities, rather, they show a correlation between the SFR and the stellar mass, i.e., galaxies with larger SFRs have  larger stellar masses. In order to test this further, we split the sample in each mass bin into two groups by the median SFR. In each mass bin and SFR group, the best-fit model spectra obtained by \textit{method1} are combined. The results are presented in Figure \ref{MZR_IntSFR} as filled and open stars for higher and lower SFR groups, respectively, as well as in Table \ref{Tab2}. For both SFRs derived from H$\alpha$ and UV, the resulting metallicities in higher SFR groups are lower than those in lower SFR groups in all mass bins, though the difference at the most massive bin in the right panel is very small. At $M_{*}=10^{10}M_{\odot}$ in the Figure \ref{MZR_IntSFR}, the gradient of metallicity against SFR from H$\alpha$ is $\Delta(12+\textrm{log(O/H)})/\Delta(\textrm{log(SFR)})\sim-0.6$ dex, which is smaller (steeper) by a factor of $\sim2$ than that of $\sim-0.3$ dex at $z\sim0.1$ \citep{Mannucci:2010p8026}. The metallicity gradient against SFR from UV is $\sim-0.4$ dex, which is almost similar to that at $z\sim0.1$.

Figure \ref{MZR_sSFR} shows that the dependence of specific SFR, derived from H$\alpha$ (left panel) and the rest-frame UV (right panel), on the mass-metallicity relation is similar to that of the SFR (Figure \ref{MZR_IntSFR}), but the trend is relatively unclear. This is probably because the correlation between the stellar mass and the SFR cancels out the dependence on the metallicity.

\begin{figure}
\begin{center}
\includegraphics[scale=0.5,angle=270]{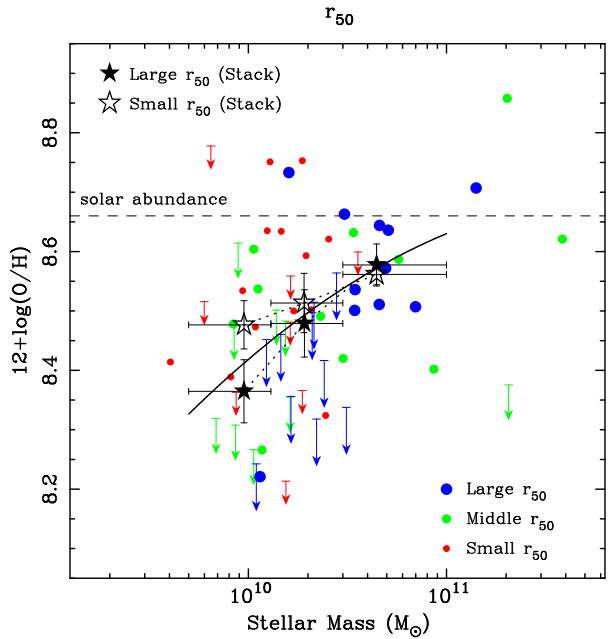}
\caption{Dependence of half light radius ($r_{50}$) on the mass-metallicity relation. The sample is divided into three groups by their half light radii with $r_{50}$ $<$ 4.38 kpc (\textit{small red}), 4.38 kpc $<$ $r_{50}$ $<$ 5.30 kpc (\textit{middle green}), $r_{50}$ $>$ 5.30kpc (\textit{large blue}). The \textit{filled stars} show the stacked results in the larger sSFR ($r_{50}$) groups, while the \textit{open stars} show those in the smaller $r_{50}$ groups in each mass bin. \textit{Solid curves} show the mass-metallicity relations described by Eq. \ref{Eq_MZR1} in Section \ref{Sec_MZR}.\label{MZR_R50}}
\end{center}
\end{figure}

We also examine the dependency of galaxy size (half light radius; $r_{50}$) described in Section \ref{Sec_ApEff} on the mass-metallicity relation. In Figure \ref{MZR_R50} and Table \ref{Tab3}, in the same manner as the SFR, we present the metallicity as a function of stellar mass dividing the sample by the size. Both individual and the results from stacking show that galaxies with larger sizes tend to have lower metallicities. In the stacking analysis, the difference is larger as the stellar mass decreases. At $M_{*}=10^{10}M_{\odot}$, the gradient of metallicity with $r_{50}$ is $\Delta(12+\textrm{log(O/H)})/\Delta(r_{50})\sim-0.1$ dex kpc$^{-1}$, which is comparable to $\sim-0.1$ dex kpc$^{-1}$ obtained at $z\sim0.1$ by \citet{Ellison:2008p7997}. Since our survey covers the CANDELS field where deep HST images are available, we can also consider the morphological information for a part of our sample. The number of such galaxies, however, is too small to be considered statistically. The relation between the metallicity and morphology will be discussed in the future works.

We further examine the dependency of other physical quantities, such as colors, surface mass density, color excess. No clear trend, however, can be seen. It is not certain whether there are truly no trends for such parameters because the sample size is still not large enough. A larger sample from the further survey with FMOS in the future is desirable in order to study these questions.

\subsection{Effects of the Sample Selection on the Stellar-Mass Relation\label{Sec_Bias}}
As mentioned in Section \ref{Sec_Selection}, our targeted sample is somewhat biased toward the larger SFR end of the galaxy sequence in the $M_{*}-SFR$ diagram as compared to the primary sample. According to the dependency of SFR on the mass-metallicity relation at $z\sim1.4$ described above, this selection may lead to a bias toward the smaller metallicity. The expected metallicity of the primary sample at a given stellar mass is examined by assuming the dependency of the SFR on the mass-metallicity relation described in the previous section. The metallicity is expressed as a linear function of intrinsic SFR from UV in each mass bin and expected metallicities of primary sample are estimated from their SFRs. The average metallicity weighted with the H$\alpha$ flux is still close to the metallicity derived in this study within $\sim0.02$ dex. 

Our sample is also slightly biased toward a larger size as compared with the primary sample. We also examine the effects on the metallicity by assuming the dependency of size on the mass-metallicity relation in the same manner described above. The resulting metallicity is also close to the observed metallicities within $\sim0.03$ dex. These effects of both SFR and size on the metallicity are smaller than the other uncertainties. Hence, we conclude that the effects of the selection bias on the mass-metallicity relation are relatively small.

The color excess of our sample distributed widely from 0.0 mag to $\sim0.6$ mag with median value of $\sim0.23$ mag. The sample is slightly biased toward a smaller color excess as compared with the primary sample (from 0.0 mag to $\sim1.0$ mag with the median value is $~0.26$ mag). However, since there is no clear dependence of color excess on the mass-metallicity relation, the selection effect on the mass-metallicity relation is small.

\begin{table}
 \begin{center}
 \caption{Metallicity from the stacked spectra in three stellar mass bins and two $r_{50}$ groups}\label{Tab3}
 \begin{tabular}{cccc}
 \hline \hline
log($M_{*}$/$M_{\odot}$) & $r_{50}^{\textrm{med}}$ (kpc) & \multicolumn{2}{c}{12+log(O/H)}\\
\cline{3-4}
 & & Small $r_{50}$ & Large $r_{50}$\\
\hline
9.98$_{-0.28}^{+0.13}$   & 4.4 & 8.477$\pm$0.041 & 8.365$\pm$0.053\\
10.28$_{-0.17}^{+0.20}$ & 4.7 & 8.513$\pm$0.050 & 8.479$\pm$0.057\\
10.65$_{-0.17}^{+0.35}$ & 5.6 & 8.562$\pm$0.017 & 8.577$\pm$0.035 \\
\hline
\end{tabular}
\end{center}
\end{table}

\section{Discussions}
\subsection{Consecutive Evolution of the Mass-Metallicity Relation from $z\sim3$ to $z\sim0$ \label{Sec_MZREvolution}}
\begin{figure}
\begin{center}
\includegraphics[scale=0.5,angle=270]{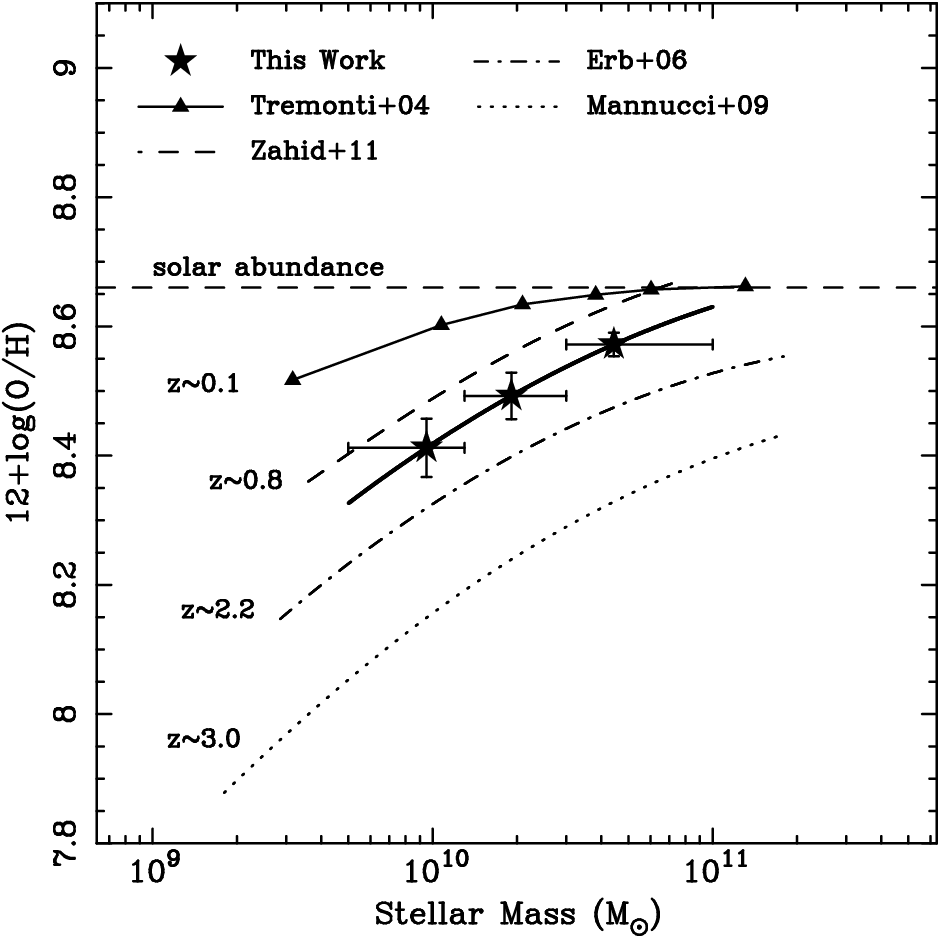}
\caption{Comparisons to the mass-metallicity relations at $z\sim0.1$ (\textit{thin solid}) by \citet{Tremonti:2004p4119}, $z\sim0.8$ (\textit{dashed}) by \citet{Zahid:2011p11939}, $z\sim2.2$ (\textit{dotted-dashed}) by \citet{Erb:2006p4143}, and $z\sim3.1$ (\textit{dotted}) by \citet{Mannucci:2009p8028}. The ranges of the relations show the observed range of the stellar masses. The stellar masses and the metallicities of other samples are converted so that the IMF and metallicity calibration are consistent with those we adopted. Horizontal dotted line indicates solar metallicity.\label{MZR_Others}}
\end{center}
\end{figure}

In Figure \ref{MZR_Others}, the mass-metallicity relation obtained in this study is shown with those at $z\sim0.1$ \citep{Tremonti:2004p4119}, at $z\sim0.8$ \citep{Zahid:2011p11939}, at $z\sim2.2$ \citep{Erb:2006p4143}, and at $z\sim3.1$ \citep{Mannucci:2009p8028}. Note again that the stellar mass and the metallicity of these samples are converted so that the IMF and metallicity calibration are consistent with those we adopted. For the conversion of the metallicity, we use prescriptions by \citet{Kewley:2008p5141} for the results at $z\sim0.8$ and by \citet{Nagao:2006p8780} for $z\sim3.1$.

In Figure \ref{MZR_Others}, it can be seen that our result at $z\sim1.4$ is located between that at $z\sim0.8$ and $z\sim2.2$ and the mass-metallicity relation evolves consecutively from $z\sim3$ to $z\sim0.1$. There appears to be no clear downsizing evolution from $z\sim3$ to $z\sim0.8$. It is interesting to note that there seems to be a downsizing-like evolution from $z=0.8-1.4$ to $z\sim0.1$. The metallicity difference between $z\sim1.4$ and $z\sim0.1$ is $\sim0.2$ dex at $M_{*}\sim10^{10} M_{\odot}$, $\sim0.1$ dex at $M_{*}\sim10^{10.5} M_{\odot}$, and almost comparable at $M_{*}\sim10^{11} M_{\odot}$ if we extrapolate our mass-metallicity relation (Eq. \ref{Eq_MZR1}). However, here we recall the saturation effect of the N2 indicator. Actually, the recalculated result of SDSS galaxies from \citet{Erb:2006p4143} shows a flattened MZ-relation at $M_{*}\gtrsim1\times10^{11}M_{\odot}$ for near-solar metallicity. Since the metallicities of our sample at $z\sim1.4$ are mostly sub-solar, they are not expected to suffer from severe saturation effects. The flat feature at the high mass end at $z\sim0.1$, however, may be due to the saturation effect. 

\begin{figure}
\begin{center}
\includegraphics[scale=0.5,angle=270]{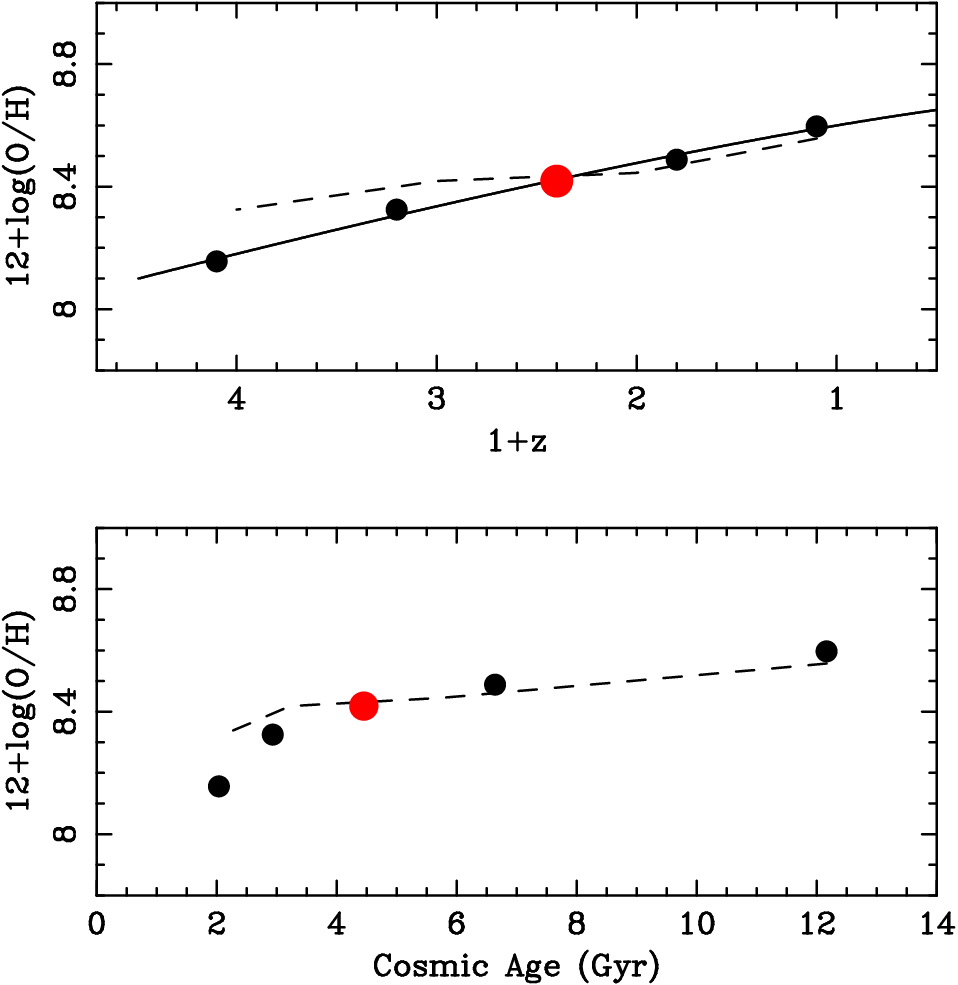}
\caption{Evolution of the mean metallicity at $M_{*}=10^{10}M_{\odot}$ against redshift (\textit{upper}) and cosmic age (\textit{lower}). The metallicities at $z\sim0.1$ \citep{Tremonti:2004p4119}, $z\sim0.8$ \citep{Zahid:2011p11939}, $z\sim1.4$ (this work), $z\sim2.2$ \citep{Erb:2006p4143}, and $z\sim3.1$ \citep{Mannucci:2009p8028} are used. In the upper panel, the linear regression line is also presented as a \textit{solid line}: $12+\textrm{log(O/H)}=8.69 - 0.086 (1+z)^{1.3}$. In both panels, the \textit{vzw} models by \citet{Dave:2011p11823} are also shown as \textit{dashed lines}. \label{CME}}
\end{center}
\end{figure}

Since the mass metallicity relation evolves without changing its shape so much except for the massive part at $z\sim0.1$, we trace the chemical evolution at $M_{*}=10^{10}M_{\odot}$. In the upper panel of Figure \ref{CME}, the metallicity at $M_{*}=10^{10}M_{\odot}$ from the mass-metallicity relation at each redshift is plotted as a function of redshift. It clearly shows that the mean metallicity increases with decreasing redshift. The cosmological evolution is well reproduced  with $12+\textrm{log(O/H)}=8.69 - 0.086 (1+z)^{1.3}$. In the lower panel, the metallicity is also plotted against the cosmic age. The metallicity evolution rate changes rapidly at $z\gtrsim2$ and seems to be saturated after that. Interestingly, the end of the rapid growth of metallicity roughly corresponds to the peak epoch of the cosmic star-formation history.
It is worth noting here that the sample selections for these mass-metallicity relations are not necessarily the same. Although the mass-metallicity relation may be affected by the sample selection, pursuing the effect at high redshift is still challenging and will be studied in the future works.

\begin{figure}
\begin{center}
\includegraphics[scale=0.5,angle=270]{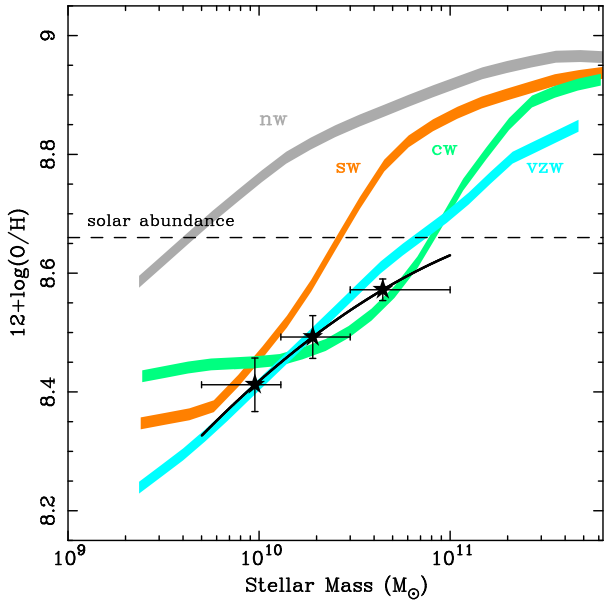}
\caption{Comparison to the theoretical models by \citet{Dave:2011p11823}. The theoretical predictions for $z\sim1.4$ are obtained by averaging those for $z=1.0$ and $z=2.0$. No winds (nw), Constant winds (cw), Slow winds (sw), and Momentum-conserving winds (vzw) are indicated by \textit{gray}, \textit{green}, \textit{orange}, and \textit{cyan}, respectively. The observed mass-metallicity relation is the same as that in the left panel of Figure \ref{MZR}.\label{MZR_Dave}}
\end{center}
\end{figure}

The mass-metallicity relation of our sample is also compared to that obtained from recent cosmological hydrodynamic simulations by \citet{Dave:2011p11823}. They implemented various outflow process and found that the momentum-conserving winds, in which both the outflow rate and the wind velocity depend on the velocity dispersion of the galaxy, i.e., galaxy mass, could reproduce the overall shapes of the mass-metallicity relations at $z\sim0.1$, $z\sim2$, and $z\sim3$ well (also shown in Figure \ref{CME}). In Figure \ref{MZR_Dave}, we show the mass-metallicity relation obtained at $z\sim1.4$ as well as  those in the simulations with four wind models: No winds (\textit{nw}), constant winds (\textit{cw}), slow winds (\textit{sw}), and momentum-conserving winds (\textit{vzw}). We take the average of results at $z\sim1$ and $z\sim2$ for comparison at $z\sim1.4$. Figure \ref{MZR_Dave} shows that our mass-metallicity relation agrees well with the \textit{vzw} model, although it also agrees with the \textit{cw} model within the error bars. This agreement between our mass-metallicity relation and the theoretical \textit{vzw} or \textit{cw} models may possibly indicate the existence of strong and mass-dependent winds in galaxies at $z=1-2$.

In Figure \ref{MZR_Scatter}, the scatter of the mass-metallicity relation at $z\sim1.4$ is compared with the predictions at $z=0-3$ with the \textit{vzw} model by \citet{Dave:2011p11823}. While the model shows the evolution of the scatter from $z=1-2$ to $z=0$, our result shows no clear evolution from $z\sim1.4$ to $z\sim0.1$. The estimated scatter in this study, however, is a lower limit. If the scatter at $z\sim1.4$ is significantly larger than that shown in Figure \ref{MZR_Scatter}, the scatter was larger at the redshift than that seen at $z\sim0.1$; the scatter of the mass-metallicity relation may decrease as the redshift decreases.  This trend is opposite to the simulation by  \citet{Dave:2011p11823}.   Further deep observations with a larger sample are desirable.

\subsection{A Possible Scenario for the Dependence of Various Parameters on the Mass-Metallicity Relation\label{Sec_Discussion2}}
\begin{figure}
\begin{center}
\includegraphics[scale=0.5,angle=270]{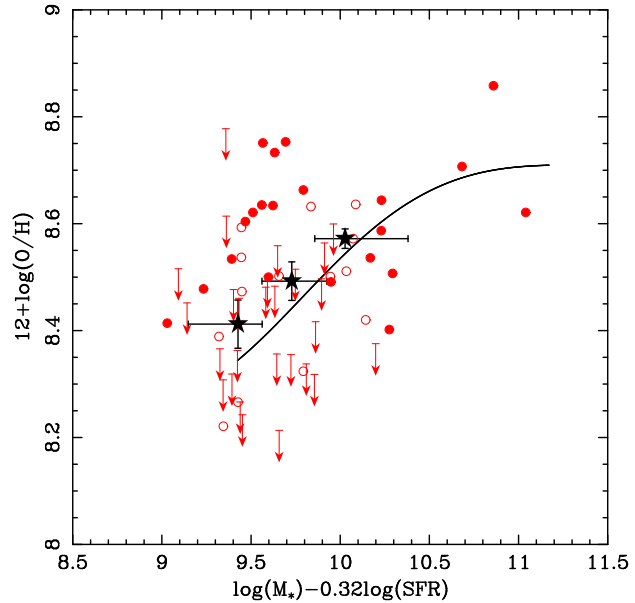}
\caption{The metallicity distribution of our sample against $\textrm{log}(\textrm{M}_{*})-0.32\textrm{log}(\textrm{SFR})$, i.e., the fundamental metallicity relation (\textit{solid line}) proposed by \citet{Mannucci:2010p8026}. Data points are based on those in the left panel of Figure \ref{MZR}.\label{FMR}}
\end{center}
\end{figure}

In Section \ref{Sec_Dependence}, we presented the dependency of the (specific) SFR and galaxy size on the mass-metallicity relation. It is suggested that the SFR or specific SFR is a second parameter which causes the scatter of the mass-metallicity relation both observationally \citep{Ellison:2008p7997, Mannucci:2010p8026, LaraLopez:2010p16580, Yates:2011p16030} and theoretically \citep{Dave:2011p11823,Yates:2011p16030} at $z\sim0$. By using SDSS galaxies at $z\sim0.1$, \citet{Mannucci:2010p8026} found that galaxies with larger SFR tend to show lower metallicity, and the scatters are significantly reduced by their proposed relation introducing SFR as the second parameter, i.e., the fundamental metallicity relation (FMR). They also found that the mass-metallicity relations obtained at $z=0.5-2.5$ are consistent with the FMR at $z\sim0.1$; the evolution of the mass-metallicity relation from $z\sim2.5$ to $z\sim0$ may be explained by the difference of SFR of the samples. In Figure \ref{FMR}, we plot our data points  on the fundamental metallicity relation by \citet{Mannucci:2010p8026}, i.e., metallicity against log($M_{*}$)-$\mu$log(SFR), where $\mu$ is a projection parameter. \citet{Mannucci:2010p8026} found that $\mu=0.32$ gives the minimum intrinsic dispersion of metallicity. The metallicity scatter of our sample is not reduced radically by the projection. The average metallicities in all mass bins  are close to the FMR, but are shifted slightly; $\mu=0.22$ gives the best fit. It should be noted, however, that SFRs in our sample are mostly larger than those in the FMR at $z\sim0.1$.

Various interpretations for the large scatter of the mass-metallicity relation and the dependence of second parameters have been discussed. One of the possible interpretation is the dilution effect of infalling gas. The infall of the gas from IGM through cold (smooth) accretion or galaxy mergers is considered to be important in galaxy growth (e.g., \cite{Bouche:2010p12762,Hopkins:2006p761}). When pristine gas falls into a galaxy, preexisting enriched gas is diluted by the fresh gas, and at the same time, the new star-formation is likely to occur, raising the SFR.

The size dependence can also be explained by the infall of gas. It is suggested that a galaxy merger event not only can supply fresh gas but also may be able to increase the effective radius of the galaxy \citep{Bezanson:2009p548,vanDokkum:2010p96}. A correlation between the signal of merger events and the metallicity has been suggested (e.g., \cite{Ellison:2008p8001}; \cite{Rupke:2010p6512}; \cite{SolAlonso:2010}). According to their results, interacting or merging galaxies tend to show lower metallicity than isolated galaxies. In our sample, we found that $\sim20$ galaxies are closely paired according to a pairing analysis (e.g., \cite{Patton:2002p868}) with requirements that (1) the projected separation is $<30$ kpc (2) the redshift separation is $<0.15$, which corresponds to $3\sigma$ of photo-$z$ uncertainty. For these objects, however, no clear trends that the interacting galaxies tend to be metal-poor can be found.

\section{Conclusions and Summary}
We present the first results obtained from near-infrared spectroscopic observations of star-forming galaxies at $z\sim1.4$ with FMOS on the Subaru Telescope. We observed K-band selected galaxies in the SXDS/UDS fields with $K\leq23.9$ mag, $1.2\leq z_{ph} \leq 1.6$, $M_{*}\geq 10^{9.5} M_{\odot}$, and expected F(H$\alpha$)$\geq10^{-16}$ erg s$^{-1}$ cm$^{-2}$. H$\alpha$ emission lines of 71 objects are detected significantly. For these objects, excluding possible AGNs identified from the BPT diagram, gas-phase metallicities and upper limits are obtained from [N\emissiontype{II}]/H$\alpha$ line ratio by carefully considering the effects of the OH-masks. We separate the sample into three stellar mass bins and stack the spectra. We obtain a mass-metallicity relation at $z\sim1.4$ from the stacking analysis. The mass-metallicity relation is located between those at $z\sim0.8$ and $z\sim2.2$. We tried to constrain an intrinsic scatter of the mass-metallicity relation and found that the scatter is $\sim0.1$ dex and is comparable to that at $z\sim0.1$ \citep{Tremonti:2004p4119}. The scatter increases as stellar mass decreases if we take the intrinsic scatters at face-value. The scatters of the mass-metallicity relation at $z\sim1.4$, however, should be lower limits; this implies that the scatter may be larger at higher redshifts. We found that the deviation from the mass-metallicity relation depends on the SFR and the half light radius: Galaxies with higher SFR and larger half light radii show lower metallicities at a given stellar mass. One possible scenario for the trends is the infall of pristine gas accreted from IGM or through merger events. Our results do not show clear FMR as proposed by \citet{Mannucci:2010p8026}. Trends of the dependence of the SFR and the size on the mass-metallicity relation are, however, still not so clear. A larger sample from the further survey with FMOS in the future may be able to reveal clearer trends, not only the dependence of the SFR and the size but also that of other parameters. The compilation of the mass-metallicity relations at $z\sim3$ to $z\sim0.1$ shows that they evolve smoothly from $z\sim3$ to $z\sim0$ without changing its shape so much (except for the massive part at $z\sim0.1$). The metallicity at $M_{*}=10^{10}M_{\odot}$ on the mass-metallicity relation increases with decreasing redshift and can be described as $12+\textrm{log(O/H)}=8.69 - 0.086 (1+z)^{1.3}$. The metallicity evolution rate was the highest at the cosmic age of $\lesssim$ 3 Gyr which was before redshift of 2. However, the result may be influenced by sample selection and further studies with large samples at high redshifts are desirable.

\bigskip
We would like to thank an anonymous referee for useful comments. We are grateful to the FMOS support astronomer Kentaro Aoki for his support during the observations. We also appreciate Soh Ikarashi, Kotaro Kohno, Kenta Matsuoka, and Tohru Nagao sharing fibers in their FMOS observations. KY is financially supported by a Research Fellowship of the Japan Society for the Promotion of Science for Young Scientists. KO's activity is supported by the grant-in-aid for Scientific Research on Priority Areas (19047003).  We acknowledge support for the FMOS instrument development from the UK Science and Technology Facilities Council (STFC). DB and ECL acknowledge support from STFC studentships. We would like to express our acknowledgment to the indigenous Hawaiian people for their understanding of the significant role of the summit of Mauna Kea in astronomical research.

\clearpage

\end{document}